\documentclass[prb,superscriptaddress,floatfix,letterpaper,twocolumn,aps,showpacs]{revtex4}
\usepackage[utf8]{inputenc}
\usepackage{natbib}
\usepackage{amsmath}
\usepackage{amsbsy}
\usepackage{amssymb}
\usepackage{scrextend}
\usepackage{graphicx}
\usepackage{color}

\newcommand{\hc}{H_{s}}
\newcommand{\hp}{H_{p}}
\newcommand{\heq}{H_{\mathrm{eq}}}
\newcommand{\hedge}{H_{\mathrm{edge}}}

\renewcommand{\a}{b}
\newcommand{\bz}{b_{0}}
\newcommand{\bone}{b_{1}}
\newcommand{\btwo}{b_{2}}
\newcommand{\ez}{e_{0}}
\newcommand{\eo}{e_{1}}
\newcommand{\et}{e_{2}}
\newcommand{\bv}{b_{\scriptscriptstyle{\triangle}}}
\newcommand{\txi}{\tilde{\xi}}

\newcommand{\fig}{Fig.\ }
\newcommand{\eq}{Eq.\ }
\newcommand{\eqs}{Eqs.\ }
\newcommand{\sect}{Sec.\ }
\newcommand{\cf}{\emph{cf.\ }}
\newcommand{\sbonlinecite}[1]{[\onlinecite{#1}]}

\DeclareMathOperator{\im}{Im}
\DeclareMathOperator{\re}{Re}
\DeclareMathOperator{\E}{E}
\DeclareMathOperator{\K}{K}
\begin{document}
\title{Suppression of Geometric Barrier in Type II Superconducting Strips}
\author{R.\ Willa}
\author{V.B.\ Geshkenbein}
\author{G.\ Blatter}
\affiliation{Institute for Theoretical Physics, ETH Zurich, 8093 Zurich,
Switzerland}
\date{\today}

\begin{abstract}
We study the magnetic response of a superconducting double strip, \emph{i.e.},
two parallel coplanar thin strips of width $2w$, thickness $d \ll w$ and of
infinite length, separated by a gap of width $2s$ and subject to a
perpendicular magnetic field $H$.  The magnetic properties of this system are
governed by the presence of a geometric energy barrier for vortex penetration
which we investigate as a function of applied field $H$ and gap parameter $s$.
The new results deal with the case of a narrow gap $s \ll w$, where the field
penetration from the inner edges is facilitated by large flux focusing.  Upon
reducing the gap width $2s$, we observe a considerable rearrangement of the
screening currents, leading to a strong reduction of the penetration field and
the overall magnetization loop, with a suppression factor reaching $\sim
(d/w)^{1/2}$ as the gap drops below the sample thickness, $2s < d$.  We
compare our results with similar systems of different shapes (elliptic,
rectangular platelet) and include effects of surface barriers as well.
Furthermore, we verify that corrections arising from the magnetic response of
the Shubnikov phase in the penetrated state are small and can be omitted.
Extending the analysis to multiple strips, we determine the specific sequence
of flux penetrations into the different strips.  Our studies are relevant for
the understanding of platelet shaped samples with cracks or the penetration
into layered superconductors at oblique magnetic fields.
\end{abstract}


\maketitle

\section{Introduction}\label{sec:introduction}
\begin{figure}[t]
\includegraphics[width=7cm]{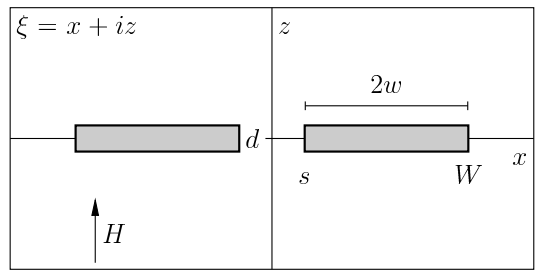}
\caption{Side-view representation ($xz$-plane) of two flat superconducting
strips (parallel to $y$) subject to a perpendicular magnetic field $H$
(directed along $z$). The cross-sections of the strips have a width $2w$ and a
thickness $d \ll w$, while the separation $2s$ between their inner edges
measures the width of the gap. The outer edges of the strips at $\pm (2w + s)$
are denoted by $\pm W$. Any position in the $xz$-plane is described by the
complex coordinate $\xi = x + i z$.}
\label{fig:sketch-double-strip}
\end{figure}

The characteristic properties of a superconductor are its diamagnetic response
\cite{Meissner_33} $M$ to an external magnetic field $H$ and its ability to
transport electric current without dissipation\cite{Onnes_11}. In the Meissner
phase the magnetic induction $B = H + 4 \pi M$ vanishes inside the
superconductor and the linear response $M = -H/4\pi$ is that of a perfect
(bulk) diamagnet. In type II superconductors, a sufficiently large magnetic
field $H>\hp$ penetrates the material via quantized flux lines (with flux
$\Phi_{0}=hc/2e$); we denote with $\hp$ the field of first penetration.
Within the mixed (or Shubnikov\cite{Shubnikov_37}) phase the presence of
vortices reduces the bulk diamagnetic signal and the magnetization $M(H)$
decreases in magnitude. The magnetic properties of the material then depend on
the behavior of the vortex state. In this paper, we determine the magnetic
response of superconducting samples of more complex shape, in particular a
double strip, two parallel coplanar thin strips of infinite length and subject
to a perpendicular magnetic field $H$, see \fig \ref{fig:sketch-double-strip}.
The response of such a system is hysteretic and dominated by the so-called
geometrical barrier\cite{Zeldov_94,Benkraouda_96}, i.e., an energy barrier
retarding the magnetic field penetration.  Our main result is an apparent
suppression of the geometrical barrier for the situation where the two strips
are closeby, i.e., separated by a narrow gap or crack. Such a suppression of
geometrical barriers may be of practical interest in experiments, as has been
the case in disentangling the vortex lattice melting- and irreversibility
lines in layered BiSCCO superconductors \cite{Majer_95} or in separating apart
the phenomenon of bulk vortex pinning by defects. So far, the geometrical
barrier has been deliberately suppressed by polishing the sample into the
shape of a prism \cite{Majer_95}; the suppression of the geometrical barrier
observed when tilting the magnetic field applied to the sample \cite{Segev_11}
and attributed to the appearance of Josephson vortex stacks resembles the
mechanism reported in the present paper.

The precise shape of the magnetization curve depends on the specific
configuration assumed by the vortices after penetration, which is determined
by the sample shape and its surface properties (we assume a sample free of
defects).  The sample surface is relevant in the determination of the
penetration field $\hp$ as defined in the asymptotic region far away from the
sample.  A flat surface parallel to the field generates an image vortex which
results in a surface barrier hindering vortices from entering the sample
\cite{Bean_64,Clem_74}. The metastable Meissner state survives until the local
field at the surface is increased beyond the critical value $\hc$ which is of
the order of the thermodynamic critical field $H_c$, $\hc \sim H_c > H_{c1}$,
with $H_{c1}$ the lower critical field. For a non-ideal surface the effective
surface barrier is reduced and assumes a value $\hc$ between $H_{c1}$ and
$H_c$.

\begin{figure}[t]
\includegraphics[width=0.45\textwidth]{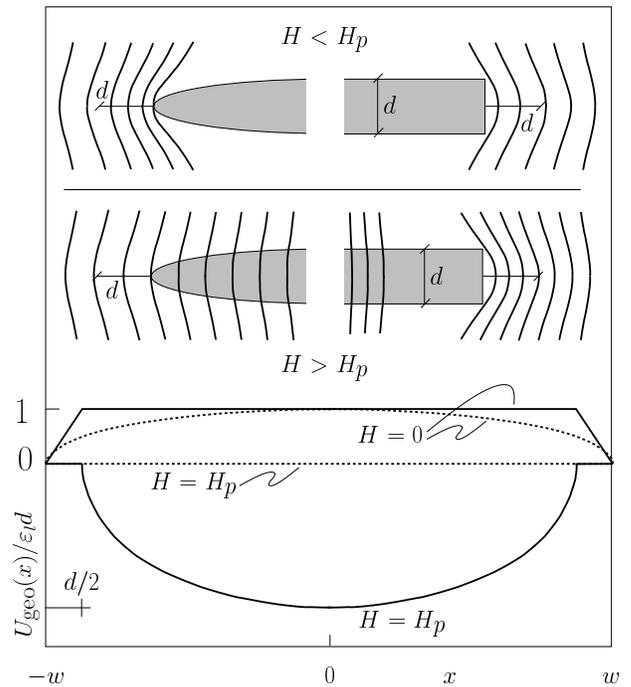}
\caption{Top: sketch of field enhancement near the edges of an elliptic-
(left) and a rectangular- (right) shaped sample. Below penetration $H < \hp$,
for both geometries, the field is enhanced by the factor $\sim \sqrt{w/d}$ a
distance $d$ away from the edges. The field remains unchanged on approaching
the rectangular edge but increases by a further factor $\sim \sqrt{w/d}$ for
the elliptic geometry.  Upon increasing $H$ beyond $\hp$, the field penetrates
homogeneously into the elliptic shaped sample and concentrates in a central
dome for the rectangular sample. This is due to the different potential
landscapes $U_{\mathrm{geo}}(x)$ (see bottom sketch) felt by the vortices
penetrating the sample at $H \sim \hp$, flat for the ellipse (dotted line) and
attractive for the rectangle (solid line). Note that the penetration fields
differ by the factor $\sqrt{d/w}$ for the elliptic and the rectangular sample.
The sketch illustrates the situation without additional surface barrier.}
\label{fig:geometries}
\end{figure}

The sample shape is relevant, too, in the determination of the penetration
field $\hp$. This is well known for elliptic-shaped samples, cf.\ Fig.\
\ref{fig:geometries}, where the magnetic field is enhanced near the sample
edge: for a cylindrical shaped diamagnetic (i.e., $\mu = 0$) sample
with an elliptic cross section of height $d$ and width $2w$, the
demagnetization factor\cite{Osborn_45} $n= 2w / (2w+d)$ generates a field enhancement $\hedge = (1-n)^{-1} H$. Correspondingly, the penetration field is given by $\hp = (1-n)\hc = d/(2w+d) \hc$.  Once the penetration field is reached, vortices enter the sample,
reversibly in the absence of a surface barrier (i.e., if $\hc =
H_{c1}$) and irreversibly else.  Without surface barrier, the vortices
distribute homogeneously inside the sample, a result that is consistent with the
constant induction inside a magnetic ellipsoid\cite{Landau_60}. On a
microscopic level,  this corresponds to an exact matching of the energy gain
of vortex motion in the field of the screening current and the energy cost
$\varepsilon_l$ associated with the increasing vortex length upon penetration,
see\ Fig.\ \ref{fig:geometries}.

For a platelet shaped sample (of width $2w$ and thickness $d$) with a
rectangular edge, the field at the boundary is enhanced as well, although
(effectively) less than for the elliptic sample. A distance $d$ away from the
edge\cite{footnote:enhancement-ellipse}, the applied field $H$ is enhanced by
a factor $\sim (w/d)^{1/2}$, resulting in a penetration field $\hp \sim
(d/w)^{1/2} \hc$. At this field strength, the barrier for vortex entry into
the sample has vanished and vortices move to the center of the sample where
they accumulate in a dome-shaped form, cf.\ Fig.\ \ref{fig:geometries}.  Under
further increase of the external field $H$, the vortex dome grows both in
height and width until the sample is fully penetrated.  In this geometry, the
cost $\varepsilon_l d$ to create the vortex is payed right upon vortex entry
at the sample edge; beyond the edge region the energy gain in the current
field $I(x)$ is no longer balanced by the energy cost and the vortex is driven
to the sample center. Hence the field penetration into the platelet shaped
sample is irreversible even in the absence of a surface barrier, what is due
to the presence of a geometric barrier defined through the energy cost for
flux entry. It is this type of geometric barrier effects
\cite{Zeldov_94,Benkraouda_96} which is at the focus of the present paper.

Another situation arises in dirty samples where vortices are pinned onto
defects. Once the surface and geometrical barriers are overcome, the vortex
arrangement may be dominated by bulk pinning and the magnetic induction
(or magnetization) is given by a Bean profile\cite{Bean_62}. What is common
to all three cases, surface-, geometric-, and bulk pinning is the
irreversible, hysteretic behavior of the magnetization $M(H)$ with changing
external field $H$. In this paper, we concentrate on the defect-free case and
thus ignore possible modifications due to bulk pinning.

The motivation to study geometrical barriers in samples of complex shape is
manifold: Originally, the understanding of the flux penetration and vortex
lattice melting in layered high-$T_c$ superconductors necessitated a proper
analysis of the vortex state in platelet-shaped samples\cite{Zeldov_94}. On
the technological side, the structuring of current-carrying strips
\cite{Benkraouda_98,Mawatari_01} enhances their critical current as the
incorporation of slits generates geometrical barriers hindering vortex motion.
Recently, Segev \emph{et al.}\cite{Segev_11} observed a structured vortex dome
in layered $\mathrm{Bi_{2}Sr_{2}CaCu_{2}O_{8+\delta}}$ samples subject to a
tilted magnetic field. This finding can be interpreted as arising from stacks
of in-plane (Josephson) vortices reducing the superconducting order
parameter\cite{Koshelev_99} and acting as weak links for the perpendicular
field (pancake vortices\cite{Feigelman_90,Clem_91}). Our analysis of vortex
penetration into a double-strip with a narrow gap, see Fig.\
\ref{fig:sketch-double-strip}, may serve as a first step towards the
understanding of flux penetration in this geometry.

From a general perspective, the magnetic response associated with
superconducting samples can be calculated numerically. Effects of complex sample
shapes, inhomogeneous material equations, and time-dependent perturbations can
then be studied quantitatively\cite{Brandt_99}. On the other hand, analytic
approaches give more qualitative insights into the system's behavior. Earlier work
on geometrical barriers in samples with more complex shapes considered the
case of two coplanar thin strips in the Meissner phase\cite{Brojeny_02} and
the full magnetization curve for a strip-shaped sample with a
slit\cite{Mawatari_03}, i.e., two strips shunted at their ends; this
ring-type topology with circulating currents exhibits a markedly different
magnetization $M(H)$ as compared to our unshunted situation. The situation of
an unshunted double stripline in the critical state was investigated in Ref.\
\sbonlinecite{Ainbinder_03}. In our work, we go beyond these results in
various ways, including the situation where the sample thickness $d$ plays an
important role.
\begin{figure}[t]
\includegraphics[width=0.45\textwidth]{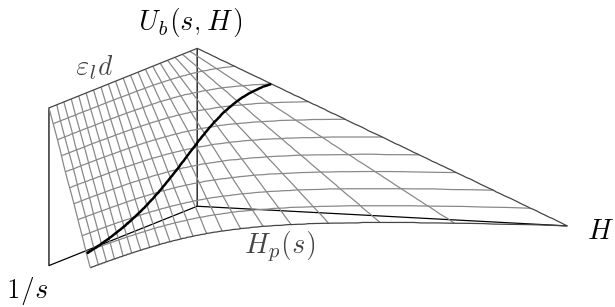}
\caption{Sketch of the geometric energy barrier $U_{b}$ for vortex penetration
as a function of the applied field $H$ and the gap parameter $s$, see also
\fig \ref{fig:sketch-double-strip}. In this Figure we neglect an additional surface barrier, i.e., $\hc = H_{c1}$. The thick black curve marks the geometric
barrier height $U_{b}^{\mathrm{eq}}(s)$ at the equilibrium field $\heq$ as
defined in \eq \eqref{eq:eq-geometric-barrier} and provides a measure for the
irreversibility of the sample. Note the rapid decrease of the geometric
barrier $U_{b}^{\mathrm{eq}}(s\ll d,H)$ with increasing field $H$ at small
separation $s$ between the two strips; the small geometric barrier
$U_{b}^{\mathrm{eq}}(s)$ tells that irreversibility is reduced when
$s~\ll~d$. Still, a finite irreversibility remains with the geometrical barrier
rapidly reinstalled when reversing the applied field.}
\label{fig:geometric-barrier}
\end{figure}%

The most pertinent new result is the dramatic suppression of the geometrical
barrier which we illustrate in \fig \ref{fig:geometric-barrier}. This
suppression is driven by a large flux-focussing into the gap between the
strips, forcing the flux penetration into the sample to start from the inner
edges. In tracing the evolution of the penetration field $\hp$ as a function
of separation $s$ between the strips, we find it decay from $\hp \sim
\sqrt{d/w} \,\hc$ at large $s$ to $\hp \sim \sqrt{sd/w^2} \log(w/s)\, \hc$ at
intermediate separation $d < s < w$ to $\hp \sim (d/w) \, \hc$ at small $s \ll
d$; the latter coincides with the result for the elliptic sample where the
geometrical barrier is absent alltogether. We emphasize, however, that the
narrow-gap double-strip still differs from the ellipse as the geometrical
barrier remains present but rapidly collapses from $\varepsilon_l d$ to zero
with increasing field, hence maintaining the hysteretic magnetization.
The latter strongly decreases with the separation $s$ between strips as well:
Within the individual strips, the penetrated field assumes a dome-like shape
which is increasingly skewed towards the gap when $s$ becomes small.
Following the change in shape of the magnetization curve through the various
regimes, we find it to shrink by a factor $\propto (s/w)^{1/2} \log(w/s)$ when
$s < w$ and by a factor $\propto (d/w)^{1/2}$ for narrow gaps $s \ll d$ when
compared to the single platelet sample; this decay of the magnetization with
decreasing $s$ ends up in a flat and nearly constant value $M = -(\hc/4\pi)
(4wd)$ at small $s \log(W/s) \ll d $.

In the following, we briefly recall the key features of the magnetic response
for elliptically shaped strips in Sec.\ \ref{sec:sec:ellipse} and proceed with
the description of coplanar parallel rectangular strips for the case
where the thickness $d$ is the smallest geometric length in the problem
(Sec.\ \ref{sec:sec:formalism}).  We review the appearance and consequences
of a geometric barrier in a single strip (Sec.\ \ref{sec:sec:single-strip})
and continue with the analysis of two adjacent strips (Sec.\
\ref{sec:sec:double-strip}) discussing the behavior of the Meissner- and
penetrated states.  In Sec.\ \ref{sec:finite-thickness}, we analyze the double
strip for the situation where the separation $2s$ between the strips is
smaller than the strip thickness $d$, $s \ll d$.  Section
\ref{sec:several-strips} is devoted to multi-strips and a summary and
conclusions are given in Sec.~\ref{sec:conclusions}.

\section{Thin Strips}\label{sec:thin-strips}
\subsection{Introduction - Elliptical strip}\label{sec:sec:ellipse}
Before considering samples with rectangular geometries, it is instructive to
revisit the magnetic properties of a flat superconducting strip with an
\emph{elliptic} cross-section. The strip extends infinitely in the
$y$-direction and the semi-axes along $x$ and $z$ are $w$ and $d/2$ ($d \ll
w$) respectively, with the upper/lower sample surface parametrized by
$z_{\pm}(x) = \pm(d/2w)\sqrt{w^2-x^2}$. The magnetic field $H$ is applied
parallel to the $z$-axis; outside the sample, $\boldsymbol{B} =
\boldsymbol{H}$, while $B_{\mathrm{el}} = \mu(B_{\mathrm{el}})
H_{\mathrm{el}}$ is constant and parallel to the $z$-axis inside the elliptic
sample\cite{Landau_60}, a consequence of the special elliptic shape. Here,
\begin{align}\label{eq:mu-from-free-energy}
   \mu(B) = \frac{B}{4\pi}\Big(\frac{dF}{dB}\Big)^{-1}
\end{align}
is the magnetic permeability of the material as obtained from the free energy
density $F(B)$. The magnetic field at the sample edges $(\pm w,0)$ is
continuous\cite{Landau_60}, $H_{\mathrm{el}} = \hedge$, where $\hedge$
denotes the magnetic field strength at the sample edge. The latter is modified
due to demagnetization effects of the sample which are described by the
geometric demagnetizing factor\cite{Osborn_45} $n = 2w / (2w + d) \approx 1 -
d/2w$. Exploiting the fact that the magnetic induction $B_\mathrm{el}$ is
constant within the ellipse, we decompose the total field
$\boldsymbol{B}(x,z)$ into two components, a constant one
$\boldsymbol{B}_\mathrm{el} = (0,0,B_\mathrm{el})$, and the remaining field
$\boldsymbol{B}_0(x,z)$ which does not penetrate the sample. Far away from the
sample, all fields point along $z$, $\boldsymbol{B}_0 \equiv (0,0,B_0^\infty)$
and we have $B_\mathrm{el} + B_0^\infty = H$.  The component
$\boldsymbol{B}_0(x,z)$ then describes the field of a perfectly diamagnetic
ellipse in the reduced external field $B_0^\infty = H-B_\mathrm{el}$. The
magnetic field at the sample edge ($x = \pm w$) points along $z$, involves the
two components $B_\mathrm{el}$ and $B_0 = B_0^\infty/(1-n)$, the latter
enhanced by demagnetization effects, and reads
\begin{align}\label{eq:edge-field-general}
   \hedge &= B_\mathrm{el} + \frac{B_0^\infty}{1-n}.
\end{align}
Using $B_0^\infty = H-B_{\mathrm{el}}$ as well as $B_{\mathrm{el}} =
\mu(B_{\mathrm{el}}) H_{\mathrm{el}} = \mu(B_{\mathrm{el}})
\hedge$, we obtain the standard formula for the
field strength inside the sample\cite{Landau_60}
\begin{align}\label{eq:elliptic-induction}
   B_{\mathrm{el}} = \frac{\mu(B_{\mathrm{el}})}{1-n[1-\mu(B_{\mathrm{el}})]}H,
\end{align}
where the value for $B_{\mathrm{el}}$ has to be determined self-consistently.
For notational simplicity we denote by $\mu$ the value for
$\mu(B_{\mathrm{el}})$ after solving the above equation.

The $\boldsymbol{B}$-field at the surface outside of the ellipse has both a
normal ($\perp$) and a tangential ($\|$) component. Their magnitudes can be
determined from the boundary conditions\cite{Landau_60}, telling that
$B_{\perp}$ and $B_{\|}/\mu$ are continuous across the surface. For the upper
surface $z=z_+(x)$ of the ellipse we find
\begin{align}\label{eq:elliptic-field-parallel-perp}
   \boldsymbol{(}B_{\|}(x),B_{\perp}(x) \boldsymbol{)} 
   &= \frac{H}{1-n(1-\mu)}\boldsymbol{(}\sin(\alpha), \mu \cos(\alpha)\boldsymbol{)},
\end{align}
where
\begin{align}
   \alpha(x) &= \arctan\bigg(\frac{d}{2w}\frac{-x}{\sqrt{w^2-x^2}}\bigg)
\end{align}
measures the angle between the external field orientation ($z$-axis) and the
direction normal to the elliptic surface at the position $\boldsymbol{(}x,
z_{+}(x)\boldsymbol{)}$. In most of the strip region (when $w-|x| \gg d^2/w$)
the surface of the ellipse is almost parallel to the $x$-axis and the above
field expression \eqref{eq:elliptic-field-parallel-perp} simplifies to
\begin{align}\label{eq:elliptic-field-parallel-perp-approx}
   \boldsymbol{(}B_{\|}(x),B_{\perp}(x) \boldsymbol{)} 
   &\approx \frac{H}{1-n(1-\mu)}\Big(\frac{-x(1-n)}{\sqrt{w^2-x^2}},\mu \Big).
\end{align}

The discontinuity of the field parallel to the boundary
determines the surface current that generates the magnetization of the sample. Using
Amp\`ere's law and defining the sheet current density $I(x) = \int_{z_-}^{z_+}
dz\, j(x,z)$ across the sample, we find
\begin{align}
   I(x)&\approx \frac{2 c}{4\pi} \big[ B_{\|}(x) - B_{\mathrm{el}} \sin(\alpha)\big]\\
    \label{eq:approx-field-at-the-surface-of-ellipse}
    &\approx -\frac{Hc}{2\pi} \frac{(1-n)(1-\mu)}{1-n(1-\mu)} 
       \frac{x}{\sqrt{w^2-x^2}}\\
       \label{eq:current-density-ellipse}
       &= -\frac{(H-B_{\mathrm{el}})c}{2\pi} \frac{x}{\sqrt{w^2-x^2}}.
\end{align}
The factor 2 originates from the two current contributions at the upper and
lower sample surface. The last expression shows that only the expelled
component $B_0^\infty = H-B_{\mathrm{el}}$ contributes to the shielding
currents.  The magnetization $M$ (per unit length) is obtained from the
relation $4\pi M / A = B_{\mathrm{el}} - H_{\mathrm{el}}$, where $A = \pi w
d/2$ is the area of the strip's cross-section. Using $H_{\mathrm{el}} = \hedge$
and \eq \eqref{eq:edge-field-general} gives for the magnetization
\begin{align}\label{eq:magnetization-ellipse-general}
   M&= -\frac{B_{0}^{\infty}}{4} w^2 
     = -\frac{H}{4} \frac{(1-n)(1-\mu)}{1-n(1-\mu)} w^2.
\end{align}
In the last equality we used $B_0^\infty = H-B_{\mathrm{el}}$ and \eq
\eqref{eq:elliptic-induction}.

\subsubsection{Meissner state}\label{sec:sec:sec:meissner-ellipse}
At low fields, the superconducting elliptic strip remains in the Meissner
state ($\mu = 0$), resulting in a vanishing induction, i.e., $B_\mathrm{el} =
B=0$. The field strength at the edge, see \eq \eqref{eq:edge-field-general},
is enhanced by the geometric factor $1/(1-n) \approx 2w/d$ as compared to the
applied field $H$.  At the sample surface, the field is everywhere tangential
and its strength is given by $H \sin(\alpha)/(1-n)$ $(\approx -H x/
\sqrt{w^{2}- x^{2}})$ as obtained from \eqs
\eqref{eq:elliptic-field-parallel-perp} and
\eqref{eq:elliptic-field-parallel-perp-approx}. The resulting sheet current
density inside the sample is obtained from \eq
\eqref{eq:current-density-ellipse},
\begin{align}\label{eq:meissner-current-ellipse}
   I(x) &\approx -\frac{H c}{2\pi} \frac{x}{\sqrt{w^2-x^2}}.
\end{align}
The perfectly diamagnetic response [\eq
\eqref{eq:magnetization-ellipse-general} with $\mu = 0$]
\begin{align}\label{eq:magnetization-meissner-ellipse}
   M &= -\frac{H}{4}w^{2}.
\end{align}
lasts until the magnetic flux starts penetrating the superconducting sample in
the form of vortices.

To bring a vortex to the position $x$ inside the sample costs an energy
 $U_{\scriptscriptstyle{L}}(x) = \varepsilon_{l} \ell(x)$, gradually
rising with the vortex length $\ell(x) = z_{+}(x) - z_{-}(x)$ from zero at the
sample edges to $d$
in the sample center; here, the line-energy $\varepsilon_{l} = \varepsilon_{0}
\log(\lambda/\xi) = \Phi_{0}(dF/dB)|_{B=0}$ is the cost per unit length
associated with the nucleation of a single vortex in the bulk superconductor.
On the other hand, the work gained from the Lorentz force [due to the current
$I(x)$ in \eq \eqref{eq:meissner-current-ellipse}] drives the vortex
entrance.  The two energy contributions can be combined to an effective
potential landscape\cite{Likharev_71} for a single vortex
\begin{align}\label{eq:energy-profile}
   U_{\mathrm{geo}}(x) 
    &= U_{\scriptscriptstyle{L}}(x) - \frac{\Phi_{0}}{c}
         \int\limits_{w}^{x} du\,I(u).
\end{align}
In the elliptical geometry, the functional form of the driving energy due to
the current \eq \eqref{eq:meissner-current-ellipse} coincides with the
geometrical thickness $\ell(x) = d\sqrt{1-x^{2}/w^{2}}$ of the sample and the
energy profile \eqref{eq:energy-profile} reduces to
\begin{align}
   U_{\mathrm{geo}}(x) &= \varepsilon_{l}\ell(x)
     \bigg(1- \frac{H \Phi_{0}}{4\pi \varepsilon_{l}} \frac{2w}{d}\bigg).
\end{align}
The barrier then vanishes throughout the sample at the penetration field
\begin{align}\label{eq:single-penetration-field-ellipse}
    \hp &= \frac{4\pi \varepsilon_{l}}{\Phi_{0}}\frac{d}{2w} = H_{c1} \frac{d}{2w}
\end{align}
where the \emph{local} field strength at the edge reaches $H_{c1}$ and the
magnetization (per unit length) as obtained from \eq
\eqref{eq:magnetization-meissner-ellipse} amounts to
\begin{align}\label{eq:single-maximal-magnetization-ellipse}
   M_{p} = -\frac{H_{c1}}{8} wd = -\frac{H_{c1}}{4\pi} \frac{\pi w d}{2},
\end{align}
with $\pi w d/2$ the cross-section of the strip.

\subsubsection{Penetrated state}\label{sec:sec:sec:penetrated-ellipse}
Beyond the field of first penetration $\hp$, vortices homogeneously flood the
sample, and the potential landscape takes the form (we replace $B_\mathrm{el}
\to B$)
\begin{align}\label{eq:energy-profile-penetrated-state}
   U_{\mathrm{geo}}(x) &= \varepsilon_{l}(B)\frac{d}{w}\sqrt{w^{2}-x^{2}} 
                           - \frac{\Phi_{0}}{c} \int\limits_{w}^{x} du\,I(u).
\end{align}
The line energy $\varepsilon_{l}(B)$ describes the energy difference (per unit
length) between the vortex state and the homogeneous field configuration,
i.e.,
\begin{align}\label{eq:effective-line-energy}
   \varepsilon_{l}(B) 
   &= \Phi_{0}\frac{d}{d B}\bigg[F(B) -\frac{B^{2}}{8\pi}\bigg]
                       = \frac{\Phi_{0} B}{4\pi} \frac{1 -\mu(B)}{\mu(B)}
\end{align}
with $F(B)$ the free energy density of the superconducting state. The second
term on the right-hand side of \eq \eqref{eq:energy-profile-penetrated-state}
is modified as well, since only the non-penetrating (diamagnetic) part $H-B$
of the field drives the diamagnetic currents in \eq
\eqref{eq:current-density-ellipse}. The resulting state remains in equilibrium
for all $H > \hp$, i.e., $U_{\mathrm{geo}}(x) \equiv 0$, and the reversible
magnetic response follows the form in Eq.\
\eqref{eq:magnetization-ellipse-general}
\begin{align}\label{eq:magnetization-superconducting-ellipse}
   M&= -\frac{H-B}{4} w^2
\end{align}
with $B$ determined by the self-consistency equation
\eqref{eq:elliptic-induction}. A finite surface barrier as discussed further below will retard the vortex penetration and generate a hysteretic response.

\begin{figure}[t]
\includegraphics[width=.45\textwidth]{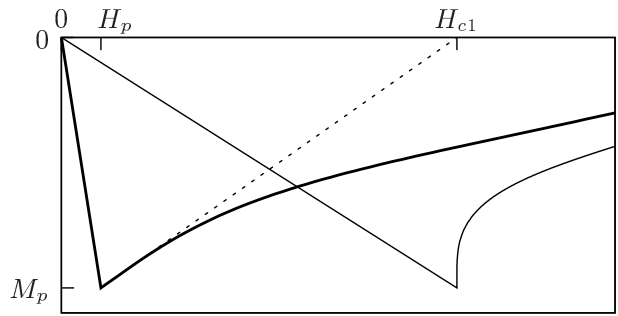}
\caption{Magnetic response of a superconducting elliptic strip with
demagnetizing factor $n \approx d/2w$ (here $n=0.9$) as obtained from \eq
\eqref{eq:magnetization-superconducting-ellipse} and with material properties
described by \eqref{eq:toy-permeability} (solid line). For comparison, we show
the bulk (Abrikosov) magnetization with the same permeability $\mu(B)$ (thin
solid line). The vertical onset in the bulk magnetization goes over into the
linear reduction of $M$ in the ellipse, extrapolating to $M=0$ at $H = H_{c1}$
(thin dashed line).}
\label{fig:elliptic-magnetization}
\end{figure}

In order to illustrate the above results, we consider a superconductor with
the Abrikosov (bulk) induction\cite{Tinkham_96}
\begin{align}\label{eq:toy-bulk-magn}
   B &= C_{1} H_{c1} \Big[\log\Big(\frac{C_{2}H_{c1}}{H-H_{c1}}\Big)\Big]^{-2}
\end{align}
near the penetration field, with $C_{1,2}$ constants of order unity. In this
equation, $H = H_\mathrm{edge}$ is the local field strength at the surface of
the bulk sample.  The magnetic permeability $\mu(B)$, can be extracted from
the above expression via the relation $\mu(B) = B/H(B)$ and we find
\begin{align}\label{eq:toy-permeability}
   \mu(B) = \frac{B}{H_{c1}} 
   \bigg[1 + C_{2}\exp\bigg(\!-\sqrt{\frac{C_{1}H_{c1}}{B}}\ \bigg)\bigg]^{-1}.
\end{align}
The linear slope $1/H_{c1}$ of the permeability near $B=0$ follows from the
vertical onset of the induction (see \eq
\eqref{eq:toy-bulk-magn}) beyond $H_{c1}$. Dropping the exponential term in \eq
\eqref{eq:toy-permeability} close to the penetration ($B \ll H_{c1}$) and
substituting $\mu$ to the self-consistency equation
\eqref{eq:elliptic-induction} we obtain the induction
\begin{align}\label{eq:ellipcit-induction}
   B(H) = (H-\hp)/n,
\end{align}
resulting in a linear decrease of the diamagnetic response,
\begin{align}\label{eq:elliptic-magnetization}
   M(H) = -\frac{\hp}{4n} w^{2}\,\Big(1-\frac{H}{H_{c1}}\Big).
\end{align}
Note that for small inductions $B \ll H_{c1}$, the diamagnetic response
\eqref{eq:elliptic-magnetization} is very different from the usual bulk
Abrikosov magnetization (see, e.g., Ref.\ \sbonlinecite{Tinkham_96}). The linear
decrease in \eq \eqref{eq:elliptic-magnetization} extrapolates to $M=0$ at
$H=H_{c1}$. The full solution of \eq \eqref{eq:elliptic-induction} for the
permeability \eqref{eq:toy-permeability} leads to the magnetic response
illustrated in \fig \ref{fig:elliptic-magnetization}.

\subsection{Rectangular strips - Formalism}\label{sec:sec:formalism}
Having familiarized ourselves with the results for the elliptic strip, we turn
our attention to strips with rectangular shape, i.e., samples with constant
thickness $d$ as opposed to the ellipse where the height is changing over the
entire sample width. Specifically, we will consider (smooth) sample edges with
a typical radius of curvature $\gtrsim d$ in contrast to the much sharper edge
of the ellipse where the radius of curvature is $d^{2}/4w \ll d$. We consider
a set of coplanar (in the $xy$-plane) and parallel superconducting strips of
infinite length (along $y$), each with a rectangular shape of width $2w$
(along $x$) and thickness $d \ll 2w$ (along $z$), subject to a perpendicular
magnetic field $H$ along $z$. The strip thickness $d$ is assumed to be the
smallest geometric length and is set to zero in the following mathematical
analysis; its finite value is properly reinstalled through appropriate
boundary conditions.  Because the system is effectively two-dimensional, we
express the magnetic field $\boldsymbol{B}(x,z)$ in the $xz$-plane through the
complex function\cite{Zeldov_94} $\mathcal{B}(\xi) = B_{z}(x,z) +
iB_{x}(x,z)$, with the two-dimensional coordinate $(x,z)$ replaced by the
complex variable $\xi = x + i z$.  The magnetostatic problem of solving the
Laplace equation ($\Delta \boldsymbol{B} = 0$) for $\boldsymbol{B}$ is
translated to a problem in complex analysis, where the holomorphic function
$\mathcal{B}(\xi)$ satisfies the Cauchy-Riemann equations (correspnding to the
magnetostatic equations $\nabla\cdot \boldsymbol{B}=0$ and $\nabla \wedge
\boldsymbol{B}=0$) in the superconductor-free region; the presence of the
superconductor is accounted for through appropriate boundary conditions. The
latter derive from two physical conditions: on the one hand, no vortices are
present in regions where current is flowing, i.e.,
\begin{align}\label{eq:bcBzI}
   B_z(x) &= 0\quad \mathrm{ when } \quad I(x) \neq 0.%
\intertext{%
Here and below we simply call `current' the sheet current density $I(x)$ flowing between $z_{\pm}=\pm
d/2$. On the other hand, no currents flow in the
vortex-filled regions,
}
   \label{eq:bcIBz}
   I(x) &= 0 \quad \mathrm{ when } \quad B_{z}(x) \neq 0.
\end{align}
This last condition
neglects the microscopic structure of the vortex state by treating the
penetrated region as magnetically inactive, $\mu = 1$; the accuracy of this
simplification will be discussed later in this section. Using Amp\`ere's law
\begin{align}\label{eq:ampere-law}
   I(x) = \frac{c}{2\pi} B_{x}(x,0^{+})
    = \frac{c}{2\pi} \mathrm{Im}[\mathcal{B}(x + i 0^{+})],
\end{align}
the boundary conditions \eqref{eq:bcBzI} and \eqref{eq:bcIBz} transform to
\begin{align}\label{eq:bcBzBx}
   B_z(x) &= 0 \quad \mathrm{ when }\quad B_{x}(x) \neq 0,\ \mathrm{and}\\
   \label{eq:bcBxBz}
   B_x(x) &= 0 \quad \mathrm{ when }\quad B_{z}(x) \neq 0. 
\end{align} 

For a single strip centered a the origin ($\xi = 0$) the holomorphic field
\begin{align}\label{eq:ant-single-strip-function}
   \mathcal{B}(\xi) &= H \sqrt{\frac{\xi^{2}-\bz^{2}(H)}{\xi^{2}-w^{2}}}
\end{align}
is known to satisfy all the above requirements\cite{Zeldov_94}; the parameter
$\bz$ then determines the field configuration of the entire system. For the
double strip studied below, the corresponding expression reads
\begin{align}\label{eq:ant-double-strip-function}
   \mathcal{B}(\xi) &=
      H \sqrt{\frac{[\xi^{2}-\bone^{2}(H)][\xi^{2}-\btwo^{2}(H)]}
      {(\xi^{2}-s^{2})(\xi^{2}-W^{2})}}.
\end{align}
Here, the strips are arranged symmetrically, extending between $\pm s$ and
$\pm W$ (with $W=s+2w$) on the $x$-axis.

In order to specify the field and current distributions for these geometries,
the parameters $\bz$, $\bone$, and $\btwo$ (with $0\leq\bz<w$ and $s < \bone \leq
\btwo < W$) describing the boundaries of the field-filled region have to be
determined from two physical conditions: First, the net current along each
strip vanishes, i.e.,
\begin{align}\label{eq:current-neutrality-condition-general}
   \int_{\mathrm{strip}}\!\!\!\!\!dx\,I(x) &= 0.
\end{align}
This (first) condition is independent of the magnetic state of the strips,
Meissner or Shubnikov. The second condition regulates the penetration process
of vortices into the superconducting sample. In the Meissner phase, no field
penetrates the superconductor and the width of the vortex dome vanishes,
imposing the (second) condition 
\begin{align}\label{eq:con2}
   \left.\begin{aligned}
  \bz &= 0 &&\textrm{for the single, or}\\
   \bone &= \btwo &&\textrm{for the double}
   \end{aligned}\right.
\end{align}
strip geometry. The second condition for the penetrated state derives from the
analysis of vortex penetration at the sample edge.  We consider a smooth edge
of shape $z_{\pm}(r) = \pm \ell(r)/2$ with $r$ measured from the sample edge, rising
to $\ell = d$ within a distance $r \approx d/2$ (e.g., $\ell(r<d/2) =
\sqrt{2rd}$). The (tangential) field $\hedge$ at the surface is assumed
constant and generates a current density $j =c \hedge/4 \pi\lambda$ at the
sample boundary, with $\lambda$ denoting the London penetration depth, $\lambda \ll
d$. A simple geometrical consideration provides us with the sheet current $I(r) = 2
(c\hedge /4\pi) \sqrt{1+[\ell'(r)/2]^2}$ and using \eq
\eqref{eq:energy-profile}, we obtain the rise of the vortex energy near the
edge
\begin{align}\label{eq:Ugeo-edge-general}
   U_{\mathrm{geo}}(r) &= \varepsilon_l \ell(r)
   - \frac{\Phi_{0}\hedge}{2\pi} \int\limits_{0}^{r} 
   du\, \sqrt{1+[\ell'(u)/2]^2}.
\end{align}
For a smooth edge with radius of curvature $\gtrsim d$ we have $\ell' \gg 1$
for $r \ll d$ (consistent with a roughly constant field $\hedge$) and we can
simplify the above expression to read
\begin{align}
   U_{\mathrm{geo}}(r) &= \varepsilon_{l} \ell(r)
   \Big[1 - \frac{\Phi_{0}\hedge}{4\pi \varepsilon_{l}}\Big].
\end{align}
Hence, we find that the energy barrier for vortex entry is eliminated when the
local field strength reaches the first critical field $\hedge = H_{c1} = 4\pi
\varepsilon_{l}/\Phi_{0}$. Once the edge region of width $d$ has been
overcome, the vortices are driven to the sample center where they arrange
within the vortex dome. The vortices deep inside the sample reduce the field
at the edge and the penetration of flux is stopped when $\hedge$ drops below
$H_{c1}$. With a further increase of the external field, vortices continue to
penetrate the sample when the condition $\hedge = H_{c1}$ is satisfied again.
This stop and go criterion for vortex penetration then is the second condition
imposed on the fields in \eqs \eqref{eq:ant-single-strip-function} and
\eqref{eq:ant-double-strip-function} and determines, together with \eq
\eqref{eq:current-neutrality-condition-general}, the parameters $\bz$,
$\bone$, and $\btwo$.

The above discussion ignores the possible presence of a surface
barrier\cite{Clem_74} appearing on small length scales below $\lambda$. In the
most effective case, this barrier further retards the penetration of vortices
until the local field reaches the critical strength $\hedge \sim H_c$. In
order to deal with the general situation accounting for effects due to a
surface barrier we denote the local critical field for vortex penetration by
$\hc$ ($H_{c1} < \hc < H_{c}$). The second condition determining the fields
\eqs \eqref{eq:ant-single-strip-function} and
\eqref{eq:ant-double-strip-function} in the penetrated ($H > \hp$) state then
can be cast in the form
\begin{align}\label{eq:condition-penetration-field}
   \hedge &= \hc.
\end{align}
The above equation replaces the condition \eq \eqref{eq:con2} valid for the
Meissner phase. In the regime of very high fields, $H > \hc$, diamagnetic
screening becomes small and the field strength at the sample edge lines up
with the applied field, $\hedge \approx H$; however, this large-field limit
will not be considered below.

Finally, we comment on the precision of this second condition: The field
strengths in \eqs \eqref{eq:ant-single-strip-function} and
\eqref{eq:ant-double-strip-function} show square-root singularities near the
sample edges. The description of the spacial dependence of the field when
approaching the edges to distances smaller than $d$ then requires a detailed
analysis of the edge region. On the other hand, the typical scale for the
field strength needed for overcoming the edge region can be obtained by the
considerations presented above, once we have a proper definition for the edge
field $\hedge$ at our disposal. Below, we identify this field strength with
the field evaluated a distance $d/2$ away from the edge, $\hedge =
B_{z}(r=-d/2)$.

The surface barrier retarding the penetration of flux appears on the
small length scale between $\lambda$ (at low fields of order $H_{c1}$) and
$\xi$ (near $H_c$). On the contrary, the \emph{geometric energy barrier} is a
macroscopic object appearing on the scale $d$. We define the \emph{geometric
barrier} $U_{b}$ of a platelet sample as the maximum of \eq
\eqref{eq:Ugeo-edge-general} that is reached near $d$. The
second term in \eq \eqref{eq:Ugeo-edge-general} then reduces the geometric
barrier linearly to zero at $\hedge = H_{c1}$ and the barrier takes the
functional form
\begin{align}\label{eq:geometric-barrier}
   U_{b} = \varepsilon_{l} d \Big(1 - \frac{\hedge}{H_{c1}}\Big) 
   = \varepsilon_{l} d \Big(1 - \frac{H}{\hp}\frac{\hc}{H_{c1}}\Big)
\end{align}
where the first (second) equality expresses the barrier in terms of the local
(asymptotic) field (note that field penetration only starts when $\hedge = \hc$, where the additional surface barrier has disappeared). While the geometric barrier \eqref{eq:geometric-barrier} only vanishes when the local field
reaches $H_{c1}$, the vortex state may become thermodynamically stable
at a lower \emph{equilibrium} field $\heq$, defined as the applied field where
a global minimum of the energy profile \eq \eqref{eq:energy-profile} develops
inside the sample. For the single (double) strip, this minimum appears at
$x_{0} = 0$ ($x_{0} = \a$) and $\heq$ is determined from the condition
\begin{align}\label{eq:def-heq}
   \varepsilon_{l} d - \frac{\Phi_{0}}{c}
   \int\limits_{e}^{x_{0}} du\, I(u)\Bigg|_{H = \heq} \!\!\!\!= 0,
\end{align}
where $e$ denotes the position of the sample edge penetrated first, $e = w$
for the single strip and $e=s$ for the double strip, see Sec.\
\ref{sec:sec:double-strip}.  The geometrical barrier at the thermodynamic field
$\heq$ 
\begin{align}\label{eq:eq-geometric-barrier}
   U_{b}^{\mathrm{eq}}
        = \varepsilon_{l} d \Big(1 - \frac{\heq}{\hp}\frac{\hc}{H_{c1}}\Big)
\end{align}
then provides us with a measure for the irreversibility of the system, see Fig.\
\ref{fig:geometric-barrier}.

Having analyzed and determined the conditions determining the parameters
$\bz$, $\bone$, and $\btwo$ in the expressions
\eqref{eq:ant-single-strip-function} and \eqref{eq:ant-double-strip-function}
for the magnetic field, we now are in a position to evaluate the magnetic
response (magnetization) of the sample. For this purpose, we make use of
Amp\`ere's law and write the holomorphic field in the form (Biot-Savart, see
also Ref.\ \sbonlinecite{Mawatari_03})
\begin{align}\label{eq:biot-savart}
   \mathcal{B}(\xi) &= H - 
    \frac{2}{c}\int_{\text{strips}}\!\!\!\! du \, \frac{I(u)}{\xi - u}.
\end{align}
This field assumes the asymptotic form (we expand for $|\xi|\gg w$)
\begin{align}\label{eq:multipole-expansion}
   \mathcal{B}(\xi) &= H - 
    \frac{2}{c\, \xi^{2}}\int_{\text{strips}}\!\!\!\!\!\! 
    du\, u\, I(u) + \mathcal{O}(\xi^{-4}),
\end{align}
where we have used that the total current in each strip vanishes.  The second
term in \eq \eqref{eq:multipole-expansion} describes the field of a line of
magnetic dipoles distributed along the $y$-axis ($\xi = 0$). We thus identify
the magnetization $M$ per unit length (from here on called magnetization) with
the expression
\begin{align}\label{eq:definition-magnetization}
   M &= \frac{1}{c} \int_{\mathrm{strips}}\!\!\!\!du\,u\, I(u).
\end{align}
This result differs from the usual textbook formula\cite{Jackson_62,
Landau_60}
\begin{align}\label{eq:magnetization-landau-lifshitz}
   \mathcal{M} = \frac{1}{2c} \int d^{3}r\;\boldsymbol{r} 
                                      \times \boldsymbol{j}(\boldsymbol{r})
\end{align}
relating the total magnetic moment  $\mathcal{M}$ to its generating current
density $\boldsymbol{j}(\boldsymbol{r})$ flowing in a loop. The translation
invariant 2D result \eqref{eq:definition-magnetization} can easily be shown to
be consistent with the 3D textbook formula for a finite size ($2L$ along $y$)
strip taking also into account the currents $j_{x}(y)$ flowing near the ends
$y = \pm L$ of the strips and closing the loop.

Formally expanding the left-hand side of \eq \eqref{eq:multipole-expansion} in
$\xi^{-2}$ and comparing terms, the magnetization can be rewritten as
\begin{align}\label{eq:magnetization-through-B}
   M(H) &= -\frac{1}{2} \left.\frac{\partial
      \mathcal{B}(\xi)}{\partial (1/\xi^{2})}\right|_{\xi^{-2} \to 0}\!.
\end{align}
The magnetic responses of the single and double strip geometries [as obtained
from \eqs \eqref{eq:magnetization-through-B},
\eqref{eq:ant-single-strip-function}, and
\eqref{eq:ant-double-strip-function}] take the particularly simple form
\begin{align}\label{eq:magnetization-via-B-single}
   M(H) &= -\frac{H}{4} (w^{2}-\bz^{2}),\\
   \label{eq:magnetization-via-B-double}
   M(H) &= -\frac{H}{4} (W^{2} + s^{2} - \bone^{2} - \btwo^{2}).
\end{align}
\subsection{Single strip}\label{sec:sec:single-strip}
We briefly review the physics of geometrical barriers for a single strip
derived by Zeldov and co-workers\cite{Zeldov_94}. The function
$\mathcal{B}(\xi)$, holomorphic in the superconductor-free region and
satisfying the required boundary conditions, is given by \eq
\eqref{eq:ant-single-strip-function}.  On the $x$-axis ($z = 0$), the magnetic
field component along $z$ is given by
\begin{align}\label{eq:single-field-general}
   B_{z}(x) &= \left\{\begin{aligned}
            &H\sqrt{\frac{\bz^{2}-x^{2}}{w^{2}-x^{2}}}
             && \mathrm{for\ } |x| \leq \bz,\\
            &H\sqrt{\frac{x^{2}-\bz^{2}}{x^{2}-w^{2}}}
             && \mathrm{for\ }w \leq |x|,\\
            &0
             && \mathrm{for\ }\bz \leq |x| \leq w.
            \end{aligned}\right.
\end{align}
The region $|x| \leq \bz$ describes the field-penetrated part of the sample
where $B_z$ is finite.  The current $I(x)$ flows in the complementary regions
$\bz \leq |x| \leq w$ inside the strip; making use of \eq
\eqref{eq:ant-single-strip-function} and Ampere's law in the form of \eq
\eqref{eq:ampere-law} we obtain the current
\begin{align}\label{eq:single-current-general}
   I(x) &= -\frac{H c}{2\pi} \frac{x}{|x|}
            \sqrt{\frac{x^{2}-\bz^{2}}{w^{2}-x^{2}}}.
\end{align}
The anti-symmetry of $I(x)$ guarantees the vanishing of the total current
as required by \eq \eqref{eq:current-neutrality-condition-general}. The
diamagnetic response resulting from these currents can be obtained with the
formula given in \eq \eqref{eq:definition-magnetization} or directly via \eq
\eqref{eq:magnetization-via-B-single}.

\subsubsection{Meissner state}\label{sec:sec:sec:Meissner-1}
In the Meissner state the field is fully expelled from the strip, $\bz=0$, and
\eqs \eqref{eq:single-field-general} and \eqref{eq:single-current-general}
simplify to
\begin{align}\label{eq:single-field-meissner}
   B_{z}(x) &= \left\{\begin{aligned}
            &H\frac{x}{\sqrt{x^{2}-w^{2}}}
             && \mathrm{for\ }w \leq |x|,\\
            &0
             && \mathrm{for\ } |x| \leq w.,
            \end{aligned}\right.
\end{align}
and
\begin{align}\label{eq:single-current-meissner}
   I(x) &= -\frac{H c}{2\pi} 
            \frac{x}{\sqrt{w^{2}-x^{2}}},
\end{align}
respectively. This anti-symmetric current density preserves the Meissner state
and is identical to the one for the elliptic strip discussed before, see \eq
\eqref{eq:meissner-current-ellipse}.  The divergencies in \eq
\eqref{eq:single-field-meissner} at $x =\pm w$ have to be cut at the distance
$\sim d$ away from the edges and we choose the specific value $d/2$. The local
field strength at the edge (we drop corrections of higher order in $d/w$)
\begin{align}\label{eq:single-field-enhancement}
   \hedge &\equiv B_{z}(w + d/2) \simeq H\sqrt{\frac{w}{d}}
\end{align}
then is enhanced by the factor $\sqrt{w/d}$. This enhancement is
parametrically smaller as compared to the flat ellipsoid with corresponding
dimensions where the enhancement factor is $2w/d$.  The response of the
superconducting strip in the Meissner state produces the magnetization [see
\eq \eqref{eq:magnetization-via-B-single} with $\bz = 0$]
\begin{align}\label{eq:single-magnetization-meissner}
   M(H) &= - \frac{H}{4} w^{2},
\end{align}
corresponding to the expulsion of the field $H$ from a region of size $\sim
w^{2}$. Similar to the currents, the diamagnetic response is identical with
that of an elliptic sample, see \eq \eqref{eq:magnetization-meissner-ellipse}.

The Meissner state becomes unstable at $H = \hp$ as determined by the
condition \eq \eqref{eq:condition-penetration-field}; with the field
enhancement given in \eq \eqref{eq:single-field-enhancement}, we find
\begin{align}\label{eq:single-penetration-field}
   \hp & \simeq \hc \sqrt{\frac{d}{w}}
\end{align}
and the (maximum) magnetization at penetration reads
\begin{align}\label{eq:single-m-at-penetration-field}
   M_{p} &= - \frac{\hc}{4} w^2 \sqrt{\frac{d}{w}}
          = - \frac{\hp}{4} w^2.
\end{align}
As discussed above, the precise value for $\hp$ depends on the details of the
edge geometry; the latter will modify the result
\eqref{eq:single-penetration-field} by a numerical factor of order unity and
affect all further results in this section in a straightforward way.  For an
elliptic strip, the larger field enhancement near the edges causes the
penetration field \eq \eqref{eq:single-penetration-field-ellipse} to be
parametrically ($\sim\!\!  \sqrt{d/w}$) smaller than that of the platelet
sample.

Although penetration is delayed to $\hp$, a field-filled state is 
thermodynamically stable (yet inaccessible due to the geometric barrier)
beyond the equilibrium field
\begin{align}\label{eq:heq-single}
   \heq = H_{c1} \frac{d}{2w}
\end{align}
as obtained from evaluating \eq \eqref{eq:def-heq}. The geometric barrier
height [from \eq \eqref{eq:eq-geometric-barrier} with $\hc = H_{c1}$] at that
specific field amounts to
\begin{align}\label{eq:eq-barrier-single}
   U_{b}^{\mathrm{eq}} = \varepsilon_{l} d \,
   \big(1 - \sqrt{d/4w}\big).
\end{align}
\subsubsection{Penetrated state}\label{sec:sec:sec:Shubnikov-1}
Increasing the external field $H$ beyond $\hp$, vortices accumulate inside the
strip in a dome-like density distribution of width $2\bz$. The field (current)
profile along the $x$-axis ($z=0$) is given by the general form
\eqref{eq:single-field-general} [\eqref{eq:single-current-general}]. The
absence of a net current inside the strip is satisfied by symmetry, $I(-x) =
-I(x)$. The evolution
\begin{align}\label{eq:single-dome-width}
   \bz^{2}(H) &\simeq w^{2} [1-(\hp/H)^{2}]
\end{align}
of the dome width as a function of the applied field $H$ is determined by
imposing a critical field strength at the edges, i.e., by solving \eq
\eqref{eq:condition-penetration-field} for $\hedge = B_{z}(w +
d/2)$. The induction in the vortex dome takes the maximal value $(\bz/w)H$ at
the gap center. For a largely penetrated strip, $w-\bz \ll w$, the induction
is almost uniform and equal to the external field, $B(x) \approx
H$. The presence of vortices inside the superconductor reduces the
diamagnetic response, see \eq \eqref{eq:magnetization-via-B-single}
\begin{align}\label{eq:single-penetrated-magnetization}
   M(H) &\simeq - \frac{\hp^{2}}{4H} w^{2}
         = - \frac{\hc^{2}}{4H} w d .
\end{align}
The applicability of the expressions \eqref{eq:single-dome-width} and
\eqref{eq:single-penetrated-magnetization} is limited to the regime where the
screening currents flow in regions much wider than the sample thickness
($w-\bz \gg d$), a limit reached when the external field $H$ is very large, of
order $\hc$. At this point, the strip is almost uniformly penetrated by the
field with $B_{z} \approx \hc$, while the remaining screening currents flow in
a narrow region of width $\sim d$ near the edges, maintaining a diamagnetic
response
\begin{align}\label{eq:magnetization-at-full-penetration}
   M(\hc) &\approx - \frac{\hp}{4} w^{2} 
             \sqrt{\frac{d}{w}}.
\end{align}
Predictions on the system's behavior for very large applied fields $H > \hc$
require a precise knowledge of the field distribution near the sample edge, a
topic which is beyond our present analysis.

The penetration process of vortices across a geometric energy barrier in a
platelet strip features a hysteretic behavior\cite{Zeldov_94,Zeldov_94_2};
upon reduction of the external field from a maximal value $H^{\star}$, the
flux $\phi_{d}^{\star}=\phi_{d}(H^{\star})$ of vortices through the sample, where
\begin{align}\label{eq:definition-flux-through-dome}
   \phi_{d} = \int\limits_{-\bz}^{\bz} dx\, B_{z}(x),
\end{align}
is trapped unless the vortex dome boundaries reach the sample edges.
Evaluating the above flux with the field \eqref{eq:single-field-general},
we find
\begin{align}\label{eq:flux-through-dome}
   \phi_{d} = 2wH\Big[\E(\bz/w) - \frac{w^{2}-\bz^{2}}{w^{2}}\K(\bz/w)\Big],
\end{align}
with $\K$ ($\E$) the complete elliptic integral of the first (second) kind
defined according to standard textbooks on mathematical functions; e.g., see
Eqs.\ (17.2.18)-(17.3.3) of Ref.\ \sbonlinecite{Abramowitz_72}, 
\begin{align}\label{eq:K}
   \K(\kappa) &= \int_{0}^{\pi/2}
                   \frac{d\theta}{\sqrt{1-\kappa^{2}\sin\!{}^{2}(\theta)}},\\
   \label{eq:E}
   \E(\kappa) &= \int_{0}^{\pi/2}
                   d\theta \sqrt{1-\kappa^{2} \sin\!{}^{2} (\theta)}.
\end{align}
For $\kappa \ll 1$, the elliptic functions show the limiting behavior
\begin{align}\label{eq:K-small}
   \K(\kappa) &= \frac{\pi}{2}\Big[ 1 + \frac{\kappa^{2}}{4}
                 + \frac{9\kappa^{4}}{64}+ \mathcal{O}(\kappa^{6}) \Big],\\
   \label{eq:E-small}
   \E(\kappa) &= \frac{\pi}{2}\Big[ 1 - \frac{\kappa^{2}}{4}
                 - \frac{3\kappa^{4}}{64}+ \mathcal{O}(\kappa^{6}) \Big],
\end{align}
while for the opposite limit, $\kappa = \sqrt{1-\nu}$ with $\nu \ll 1$, we find 
\begin{align}\label{eq:K-large}
   \K(\sqrt{1-\nu}) &= \frac{1}{2}\log\Big(\frac{16}{\nu}\Big)
                    - \frac{\nu}{8}\Big[2 - \log\Big(\frac{16}{\nu}\Big)\Big]
                    + \mathcal{O}(\nu^{2}),\\
   \label{eq:E-large}
   \E(\sqrt{1-\nu}) &= 1 
                       - \frac{\nu}{4}\Big[1 
                                       - \log\Big(\frac{16}{\nu}\Big)\Big]
                       + \mathcal{O}(\nu^{2}).
\end{align}
The constraint $\phi_{d}(H<H^{\star}) = \phi_{d}^{\star}$ reduces
to a condition for the dome width $\bz(H)$ of the form
\begin{align}\label{eq:descending-branch-condition-single}
   \E(\bz/w) - \frac{w^{2}-\bz^{2}}{w^{2}}\K(\bz/w)
          &= \frac{\phi_{d}^{\star}}{2w\, H},
\end{align}
in agreement with Ref.\ \sbonlinecite{Benkraouda_96}. The left-hand side is
limited by unity from above (for $\bz = w$). Upon decreasing $H$, the vortex
dome expands over the sample until reaching the edge. Since for a large dome,
$w - \bz \ll w$, the induction is uniform and equal to $H$, we find that
vortices leave the sample at $H = H_{\mathrm{ex}}= \phi_{d}^{\star}/2w$ where
formally $\bz = w$ and $M=0$.
\begin{figure}[t]
\includegraphics[width=0.48\textwidth]{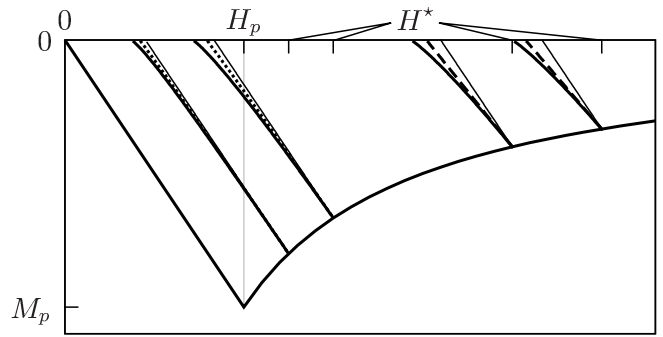}
\caption{The magnetization for the descending field branches are shown for
different values of the turning field $H^{\star}$ ($H^{\star}/\hp$ = 1.25,
1.5, 2.5, 3). The numerical solution of \eq
\eqref{eq:descending-branch-condition-single} (thick solid lines) is compared
to the magnetic response obtained from small domes (thin solid lines)
featuring a constant Meissner slope, see \eq
\eqref{eq:magnetization-descending-single-small-dome-1}. For $H^{\star}$ =
1.25$\hp$ and 1.5$\hp$, the dotted curves show the magnetic response as
obtained from a next-to-leading order expansion of \eq
\eqref{eq:descending-branch-condition-single} in $\bz/w$, see
\eq\eqref{eq:descending-branch-single-small-dome-4}. For $H^{\star}$ =
2.5$\hp$ and 3$\hp$, where the dome is sufficiently large at $H^{\star}$,
i.e., $\nu^{\star} = (\hp/H^{\star}) \ll 1$, the magnetization is well
described by the expression
\eqref{eq:analytic-returning-single-strip} shown as dashed lines.}
\label{fig:descending-branch-single}
\end{figure}%

For a narrow dome $\bz/w \ll 1$, the above condition
\eqref{eq:descending-branch-condition-single} can be simplified using the
asymptotic expressions \eqref{eq:K-small} and \eqref{eq:E-small} for the
elliptic functions; to lowest (quadratic) order in $\kappa = \bz/w$, we find
\begin{align}\label{eq:descending-branch-single-small-dome-1}
   \frac{\bz^{2}}{w^{2}} &= \frac{4}{\pi}\frac{H_{\mathrm{ex}}}{H}.
\end{align}
The growth of the dome width $\bz^{2} = \bz^{\star}{}^{2} H^{\star}/H$ [with
$\bz^{\star} = \bz(H^{\star})$] results in a magnetic response of the form
\begin{align}\label{eq:magnetization-descending-single-small-dome-1}
   M(H) &= -\frac{H}{4} w^{2} + \frac{H^{\star}}{4} {\bz^{\star} }^2
\end{align}
with a slope identical to the Meissner state. Higher order corrections
(quartic in $\kappa = \bz/w$) are straight forwardly obtained from \eqs
\eqref{eq:K-small} and \eqref{eq:E-small}: the condition
\eqref{eq:descending-branch-condition-single} then yields
\begin{align}\label{eq:descending-branch-single-small-dome-4}
   \frac{\bz^{2}}{w^{2}} &= 4 \bigg\{
            \sqrt{1 + \frac{H^{\star}}{H}
               \Big[
                  \Big(1 + \frac{1}{4}\frac{\bz^{\star}{}^{2}}{w^{2}}
                  \Big)^{2} - 1
               \Big]
            } -1
         \bigg\}.
\end{align}
Inserting this solution into the expression \eqref{eq:magnetization-via-B-single} for the
magnetization, the Meissner slope is corrected according to
\begin{align}\nonumber
   \frac{dM}{dH} &= - \frac{w^{2}}{4} \bigg\{
                     1 - \frac{1}{2}\Big(\frac{H^{\star}}{H}\Big)^{2}
                        \Big[
                           \Big(1 + \frac{1}{4}\frac{\bz^{\star}{}^{2}}{w^{2}}
                           \Big)^{2} - 1
                        \Big]^{2}
                  \bigg\}\\
            &\approx - \frac{w^{2}}{4} \Big[1- \frac{1}{8}
                      \Big(
                         \frac{H^{\star}}{H} \frac{\bz^{\star}{}^{2}}{w^{2}}
                      \Big)^{2}
                   \Big].
\end{align}

The rapid growth of the dome-width on both the field-increasing (filling the
dome with additional flux) and decreasing (expanding the dome at fixed flux)
branches leads to a fast violation of the condition $\bz \ll w$ assumed above
and hence these results have a rather limited range of validity.  Another
limit is reached when $\bz$ is large, $w - \bz \ll w$.  Defining $\nu = 1 -
\bz^{2}/w^{2}$, the asymptotic expressions \eqref{eq:K-large} and
\eqref{eq:E-large} can be used to simplify (up to linear order in $\nu$) the
condition \eqref{eq:descending-branch-condition-single} to
\begin{align}\label{eq:simplified-returning-branch}
   1 - \frac{\nu}{4} \Big[\log\Big(\frac{16}{\nu}\Big) + 1\Big]
       &= \frac{H_{\mathrm{ex}}}{H}.
\end{align}
In most of the $M$-$H$-diagram, the system's magnetic response on the descending
branch then is given by $M(H) = -H w^{2} \nu(H)/4$. Taking the derivative of $M$
with respect to $H$, the slope of the descending branch can be evaluated and,
after some reordering, we find that
\begin{align}\label{eq:slope}
   \frac{dM}{dH} &= -\frac{w^{2}}{4} \frac{4 - \nu}{\log(16/\nu)}.
\end{align}
The derivative \eqref{eq:slope} deviates from the Meissner slope $-w^{2}/4$ by
a numerical factor which assumes the value $\approx 1.01$ for $\nu \sim 1/2$,
when the previous approach of a narrow dome predicts a perfect Meissner slope
[see \eq \eqref{eq:magnetization-descending-single-small-dome-1}]. In the
regime of applicability, where $\nu$ may change by several orders of
magnitude, the factor $(4-\nu)/\log(16/\nu)$ changes noticeably but not
parametrically. Typically, the slope of the descending branch is numerically
close to the Meissner slope within the parameter range under consideration,
see Fig.~\ref{fig:descending-branch-single}. For large reversal fields
$H^{\star} \gg \hp$, we replace the parameter $\nu(H)$ by its value at the
field reversal $\nu(H^\star) = \nu^{\star} = 1 - \bz^{\star}{}^{2}/w^{2} =
(\hp/H^\star)^{2}$, where we have used \eq \eqref{eq:single-dome-width}.
The magnetization
\begin{align}\label{eq:analytic-returning-single-strip}
   M(H) = M(H^{\star}) - \frac{H-H^{\star}}{4} w^{2}
                         \frac{4 - \nu^{\star}}{\log(16/\nu^{\star})} 
\end{align}
as obtained from \eq \eqref{eq:slope} and integration from $H^{\star}$ to $H$
provides a good description of the descending branch in this regime, see \fig
\ref{fig:descending-branch-single}.

As the boundaries of the dome approach the edges of the strip to a distance
$\sim d$ (which is the case when $H \approx [1+ \mathcal{O}(d/w)]
H_{\mathrm{ex}}$) the precise geometric shape of the sample edge needs to be
taken into account, requiring a more accurate analysis going beyond the
present description.  An attempt to cope with this situation has been
undertaken by Zeldov and co-workers in Refs.\
\sbonlinecite{Zeldov_94_2,Morozov_96,Morozov_97}.
\subsubsection{Magnetization of the vortex dome}
\label{sec:sec:sec:magnetic-medium-1}
The physical properties of quantized flux lines appeared in the above analysis
merely as a criterion for vortex entry at the sample edges.  The vortex dome
in the penetrated state has been described by a smooth field $B_z(x) \neq 0$
residing in a magnetically inactive medium with $\mu = 1$ whose extend
$[-b_0,b_0]$ derives from the solution ${\cal B}(\xi)$ of the boundary value
problem. In reality, the vortex state in the dome is described by a field
$h(x)$ modulated on the scale of the inter-vortex distance due to vortex
currents. In the following, we show that the currents associated with the
vortex state in the dome generate a magnetization which remains small as
compared to the magnetization produced by the screening currents flowing in
the field-free regions.

An analogous problem appears in the context of surface barriers as discussed
by Clem\cite{Clem_74} and by Koshelev\cite{Koshelev_94}: quite similar to
our analysis, in Ref.\ \sbonlinecite{Clem_74} the vortex-penetrated bulk,
separated from the boundary by a layer of screening (Meissner) currents, has
been described by an induction $B_z$ averaged over the inter-vortex spacing.
This approximation neglects all field and current modulations due to the
vortex state and the resulting magnetization density is given by\cite{Clem_74}
\begin{align}\label{eq:magnetization-surf-barrier-Clem}
   m(H) &= -\frac{H}{4\pi} \big[1 - \sqrt{1 - (\hc/H)^{2}}\,\big].
\end{align}
A way to account for the local currents in the vortex state has been proposed
by Koshelev\cite{Koshelev_94}, who found that these contribute a paramagnetic
correction $\delta m = (\sqrt{3}/48) (\Phi_{0}/4\pi \lambda^{2})$ to the
magnetization density $m(H)$ in the limit $B \gg \Phi_{0}/\lambda^{2}$.
Following a similar ideology as in Ref.\ \sbonlinecite{Koshelev_94}, we
describe the flux-filled region in terms of a vortex lattice along $z$ with
vortex rows aligned along $y$ and separated by $\bv$ in the $x$-direction with
$\bv^{2} = (3/4)^{1/2}\Phi_{0}/B_{z}$. While in Ref.\
\sbonlinecite{Koshelev_94} $B_z(x)$ was determined self-consistently, here, we
estimate the corrections to the magnetization by adopting the averaged field
$B_z(x)$ obtained from the above analytic solution.  In our strip geometry,
the spacing $\bv$ between vortex-rows slowly varies along $x$, as the
induction $B_{z}$ changes on macroscopic length scales. The connection between
the local field $h(x)$ and the induction $B_{z}(x)$ is given by the average
\begin{align}\label{eq:average-induction}
   B_{z}(x) = \frac{1}{\bv}\int\limits_{x-\bv/2}^{x+\bv/2} dx' h(x').
\end{align}
The local field $h(x)$ satisfies the one-dimensional London equation $\lambda^{2} h''(x) +
h(x) = 0$ between the vortex rows with the boundary conditions replaced by
the constraint \eqref{eq:average-induction}. For a slowly varying dome
profile, i.e., $\bv \partial_{x} B_{z}(x) \ll B_{z}(x)$, we obtain the field
modulation between vortex rows 
\begin{align}
   h(x) &\approx B_{z}(x_{c}) \frac{\bv}{2\lambda}
           \frac{\cosh[(x-x_{c})/\lambda]}{\sinh(\bv/2\lambda)},
\end{align}
with $x_{c}$ the center between the two adjacent rows and $|x-x_{c}| < \bv/2$.
Amp\`ere's law then provides us with the current profile
\begin{align}
   j(x) &\approx - \frac{B_{z}(x_{c})c}{4\pi} \frac{\bv}{2\lambda^{2}}
                   \frac{\sinh[(x-x_{c})/\lambda]}{\sinh(\bv/2\lambda)}
\end{align}
and we can evaluate the associated average magnetization density at the
vortex location $x_v$
\begin{align}
   m(x_{v}) &\approx \frac{1}{\bv c}
            \int\limits_{x_{v}-\bv/2}^{x_{v}+\bv/2} dx' x' j(x')\\
            \label{eq:material-magnetization}
            &\approx \frac{B_{z}(x_{v})}{4\pi} 
   \Big[1 - \frac{\bv}{2\lambda} \frac{1}{\sinh(\bv/2\lambda)}\Big].
\end{align}
For small fields $B_{z} \ll H_{c1}$, we find that $m(x) \approx
B_{z}(x)/4\pi$, while the magnetization density saturates at
$(\Phi_{0}/4\pi\lambda^{2}) \sqrt{3}/48$ for large fields $B_{z} \gg H_{c1}$,
consistent with the results presented in Ref.\ \sbonlinecite{Koshelev_94}.  In
order to estimate the correction to the strips' magnetic response, we
introduce the upper bound
\begin{align}\label{eq:ub-magnetization}
   m(x) \leq \frac{B_{z}(x)}{4\pi} \frac{H_{c1}}{H_{c1}+B_{z}(x)}
\end{align}
with the correct asymptotic behavior for $B \ll H_{c1}$ and logarithmically
[$\propto \log(\lambda/\xi)$] overestimating the magnetization when $B \gg
H_{c1}$. Integrating $m(x)$ over the dome and replacing the dome profile
$B_z(x)$ by its maximum $H \bz/w$ at the center, see \eq
\eqref{eq:single-field-general}, we obtain the bound
\begin{align}\label{eq:magnetic-correction-estimated}
   \delta M < \frac{H}{4\pi} \frac{H_{c1}}{H_{c1} 
   + H \bz/w} \frac{\bz^{2}}{w^{2}}2w d.
\end{align}
For a small dome, $\bz \ll w$, this expression simplifies to
\begin{align}\label{eq:magnetic-correction-estimated-low-field}
   \delta M < \frac{H}{4\pi} \Big(1 - \frac{\hp^{2}}{H^{2}} \Big)2w d,
\end{align}
whereas for a large part of the penetrated region $d \ll w-\bz(H) \ll w$, 
we find that 
\begin{align}\label{eq:magnetic-correction-estimated-high-field}
   \delta M < \frac{H}{4\pi} \frac{H_{c1}}{H_{c1} + H} 2w d.
\end{align}
As a result, the correction $\delta M$ due to the reversible magnetization
measured on the magnetization $M$ of the screening currents \eq
\eqref{eq:single-penetrated-magnetization} is bounded from above by
\begin{align}\label{eq:relative-corrections-single}
   \frac{\delta M}{M} 
   &< \frac{2}{\pi} \frac{H^{2}}{\hc^{2}} \frac{H_{c1}}{H_{c1}+H}.
\end{align}
In the absence of a surface barrier ($\hc = H_{c1}$) these corrections are
small and become of order unity at the largest fields $H \sim H_{c1}$ where
our analysis applies.  In the presence of a large surface barrier where $\hc
\gg H_{c1}$, the corrections are even smaller and reach a maximum $\sim
H_{c1}/\hc \ll 1$ when $H \sim \hc$.  We conclude that the corrections arising
from the vortex currents can be omitted in the single strip geometry.

\subsection{Double strip}\label{sec:sec:double-strip}
We now investigate the double-strip configuration defined in \fig
\ref{fig:sketch-double-strip}, a system of two coplanar, parallel strips of
width $2w$ each and separated by a gap $2s$. Assuming a gap that is large as
compared to the strip thickness, $s \gg d$, the system can be treated within
the framework introduced in \sect \ref{sec:sec:formalism}. The holomorphic
function has been presented in \eq \eqref{eq:ant-double-strip-function}, from
which the [symmetric, $B_z(-x) = B_z(x)$] field and [anti-symmetric, $I(-x) =
-I(x)$] current distribution on the $x$-axis can be readily deduced
\begin{align}\label{eq:double-field-general}
   \frac{B_{z}(x)}{H} &= \left\{ \begin{aligned}
            &\sqrt{\frac{(\bone^{2}-x^{2})(\btwo^{2}-x^{2})}
                 {(s^{2}-x^{2})(W^{2}-x^{2})}}
             && \mathrm{for\ } 0 \leq x \leq s,\\
            &\sqrt{\frac{(x^{2}-\bone^{2})(\btwo^{2}-x^{2})}
                 {(x^{2}-s^{2})(W^{2}-x^{2})}}
             && \mathrm{for\ } \bone \leq x \leq \btwo,\\
            &\sqrt{\frac{(x^{2}-\bone^{2})(x^{2}-\btwo^{2})}
                 {(x^{2}-s^{2})(x^{2}-W^{2})}}
             && \mathrm{for\ } W \leq x,\\
            &0
             && \mathrm{otherwise},
            \end{aligned}\right.
\end{align}
and 
\begin{align}\label{eq:double-current-general}
      \frac{2\pi I(x)}{cH} &= \left\{ \begin{aligned}
            &\sqrt{\frac{(\bone^{2}-x^{2})(\btwo^{2}-x^{2})}
                 {(x^{2}-s^{2})(W^{2}-x^{2})}}
             && \mathrm{for\ } s \leq x \leq \bone,\\
            &-\sqrt{\frac{(x^{2}-\bone^{2})(x^{2}-\btwo^{2})}
                 {(x^{2}-s^{2})(W^{2}-x^{2})}}
             && \mathrm{for\ } \btwo \leq x \leq W,\\
            &0
             && \mathrm{otherwise}.
            \end{aligned}\right.
\end{align}
The resulting magnetization is given by \eq
\eqref{eq:magnetization-via-B-double}.

\subsubsection{Meissner state}\label{sec:sec:sec:Meissner-2}
\begin{figure}[t]
\includegraphics[width=0.48\textwidth]{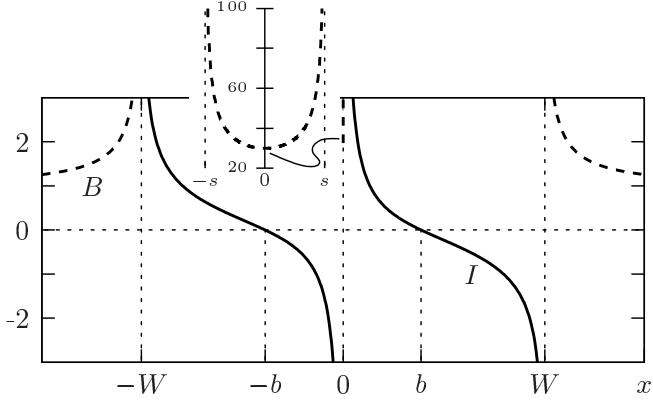}
\caption{Normalized current density $2 \pi I(x)/H c$ (solid line) flowing
along the $y$-direction and dimensionless magnetic field $B_{z}(x)/H$ (dashed
line) of a double-strip in the Meissner state. The two strips with width $2w$
and thickness $d$ ($d/w \to 0$) are separated by a gap $2s$ (here $w/s=100$).
According to \eqs \eqref{eq:ant-double-strip-function} and
\eqref{eq:double-zero-current-result}, the local current reverts its sign at
$\pm \a$, with $\a \approx 0.38 \, W$. The magnetic field inside the gap between
the strips (see inset) is far above the range of this graph, $B_{z}(|x|<s)/H
\geq \a^{2}/Ws \approx 30$.}
\label{fig:double-strip-meissner-d-ll-s}
\end{figure}

In the (low-field) Meissner state the parameters $\bone$, $\btwo$ in \eq
\eqref{eq:ant-double-strip-function} coincide, $\bone = \btwo = \a$, with $\pm \a$
marking the the positions inside the strips where the current density changes
sign (see \fig \ref{fig:double-strip-meissner-d-ll-s}). The magnetic field
component $B_{z}$ [from \eq \eqref{eq:double-field-general}] is non-vanishing
whenever $|x|\leq s$ or $W \leq |x|$ and reads
\begin{align}\label{eq:double-field-meissner}
   B_{z}(x) &=
    H \frac{|x^{2}-\a^{2}|}
                 {\sqrt{(x^{2}-s^{2})(x^{2}-W^{2})}}.
\end{align}
In the complementary region $s \leq |x| \leq W$, the screening current
\begin{align}\label{eq:double-current-meissner}
   I(x) &=  - \frac{H c}{2\pi} \frac{x}{|x|} \frac{x^{2}-\a^{2}}
                 {\sqrt{(x^{2}-s^{2})(W^{2}-x^{2})}}
\end{align}
guarantees a perfect diamagnetic (Meissner) response
\begin{align}\label{eq:double-strip-magnetization-meissner}
   M(H) &= -\frac{H}{4} (W^{2} + s^{2} - 2\a^{2}),
\end{align}
with $\a$ independent of $H$. The condition
\eqref{eq:current-neutrality-condition-general} that no net current flows
along each strip requires that
\begin{align}\label{eq:double-zero-current-cond}
   \int\limits_{s}^{W} \frac{dx\,x^{2}}{\sqrt{(x^{2}-s^{2})(W^{2}-x^{2})}} &= 
     \int\limits_{s}^{W} \frac{dx\,\a^{2}}{\sqrt{(x^{2}-s^{2})(W^{2}-x^{2})}},
\end{align}
from which we find the value of $\a$,
\begin{align}\label{eq:double-zero-current-result}
   \a^{2} &= W^{2} \frac{\E(\kappa')}{\K(\kappa')},
\end{align}
in agreement with Ref.\ \sbonlinecite{Brojeny_02}. Here, $\K$ ($\E$) is the
complete elliptic integral of the first (second) kind, as defined in \eq
\eqref{eq:K} [\eqref{eq:E}], and $\kappa' = \sqrt{1-\kappa^{2}}$ is the
complementary modulus of $\kappa = s/W$. For large gaps, the double strip
behaves as two independent strips: indeed, for $s/w \to \infty$, the parameter
$\a$ approaches the sample center $w+s$ and the magnetization assumes the
asymptotic value $M(H) \to -H w^{2}/2$, twice that of an isolated strip, see
\eq \eqref{eq:single-magnetization-meissner}.

Let us then focus on the opposite limit $s \ll W=2w+s$, where the right hand
side of \eq \eqref{eq:double-zero-current-cond} shows a logarithmic divergence
$\propto \log(W/s)$, while the left hand side is regular; in this limit, the
parameter $\a$ takes the asymptotic form
\begin{align}\label{eq:a-over-W-ratio}
   \a^{2} &= \frac{W^{2}}{\log{(4W/s)}},
\end{align}
and the position $\a$ where the current $I(x)$ changes sign is no longer at
the sample center but has shifted towards the inner edge, see Figs.\
\ref{fig:double-strip-meissner-d-ll-s}, \ref{fig:b1-n-b2} and \ref{fig:a-of-s}.  The
magnetization (per unit length) \eqref{eq:double-strip-magnetization-meissner}
to leading order in $s/W$ reads,
\begin{align}\label{eq:meiss-mag-double-strip}
   M(H) &= -\frac{H}{4}W^{2}\Big[1 - \frac{2}{\log(4W/s)}\Big].
\end{align}
In the limit $s/W \to 0$, the Meissner slope approaches that of a single strip
with double width, see \eq \eqref{eq:single-magnetization-meissner}. We
conclude that over the full range of gap widths $s$ (from $s \gg W$ down to
$s/W \to 0$) the slope in the magnetization of the Meissner state increases
only by a factor 2.

For the double strip geometry, the flux (per unit length) $\phi_{g}$ passing
through the gap $|x| < s$ is defined as the $z$-component of the magnetic
field \eqref{eq:double-field-meissner} integrated over the gap width,
\begin{align}\label{eq:flux}
   \phi_{g} &=\!\!\int\limits_{-s}^{s}\! dx\,B_{z}(x)
         = 2 W \Big[\E(\kappa)
                      - \Big(1-\frac{\a^{2}}{W^{2}}\Big)\K(\kappa)\Big] H,
\end{align}
where the elliptic functions are evaluated at $\kappa = s/W$.  In the regime
of almost independent strips, $s \gg W$, the flux $2s H$ of the homogeneous
field in the empty gap region is enhanced by half of the flux $\phi_{b} = 4w
H$ blocked by the two strips, thus adding up to $\phi_{g} \approx (2s + 2w)H$. In the opposite limit $s \ll W$, the expression \eqref{eq:flux} for the flux in the gap simplifies to
\begin{align}\label{eq:flux-t-s-l}
   \phi_{g} &\simeq \frac{\pi \a^{2}}{W} H \simeq \frac{\pi W}{\log(4W/s)} H.
\end{align}
An essential part (up to a logarithmic factor) of the blocked flux
$\phi_{b} = 2W H$ is pushed through the gap. This slow reduction of $\phi_{g}$
upon reducing $s$ goes hand in hand with an enhancement of the field strength
at the gap center
\begin{align}\label{eq:double-field-enhancement-gap-center}
   B_{z}(0) &= H \frac{\a^{2}}{s W}%
             = \frac{2}{\pi}\frac{\phi_{g}}{2s}%
             = H \frac{W/s}{\log(4W/s)}
\end{align}
and near the inner edges
\begin{align}\label{eq:double-field-enhancement}
   B_{z}(s - d/2)%
        &\simeq H\frac{\a^{2}}{\sqrt{sd}\,W}
        = H\frac{W/\sqrt{sd}}{\log(4W/s)}.
\end{align}
This last expression is parametrically larger than the enhancement observed at
the edge of an isolated strip, see \eq \eqref{eq:single-field-enhancement}.
Note that the field inside the gap is far from constant, but increases by a
factor $\sqrt{s/d}$ from the gap center to one strip edge, see inset in \fig
\ref{fig:double-strip-meissner-d-ll-s}). On the other hand, the field strength
near the outer edges
\begin{align}\label{eq:double-field-enhancement-outer}
   B_{z}(W + d/2)%
        &\simeq H \frac{W^{2}-\a^{2}}{W\sqrt{Wd}}
        = H\sqrt{\frac{W}{d}}\Big[1 - \frac{1}{\log(4W/s)}\Big] 
\end{align}
is comparable to that of an isolated strip, see \eq
\eqref{eq:single-field-enhancement}.  From this analysis we conclude that the
local critical field $\hc$ is first reached near the inner edges, such that
the penetration of vortices occurs from \emph{inside}. The field
of first penetration $\hp$ then is determined by the condition
\begin{align}\label{eq:criticality-condition-double-strip}
   \hedge&= B_{z}(s - d/2) = \hc
\end{align}
and making use of \eq \eqref{eq:double-field-enhancement} we find that the
penetration field for small gaps $s \ll W$
\begin{align}\label{eq:double-penetration-field}
   \hp &\simeq \hc \sqrt{\frac{s d}
                               {W^{2}}} \frac{W^{2}}{\a^{2}}
    = \hc \sqrt{\frac{s d}{W^{2}}} \log{(4W/s)}
\end{align}
is substantially reduced as compared to the one for isolated strips $\hp
\simeq \hc \sqrt{d/w}$. As discussed for the single strip, see \eq
\eqref{eq:single-penetration-field} and thereafter, the precise edge geometry
will alter the above expression for $\hp$ by a numerical factor of order unity
(the same factor as for the single strip), a correction that will be neglected
in the following.  At penetration $H=\hp$, the Meissner state reaches the
maximal diamagnetic response [see \eq \eqref{eq:meiss-mag-double-strip}]
\begin{align}\label{eq:max-magnetization-double-thin-strip}
   M_{p} &= -\frac{\hc}{4}W \sqrt{s d}\, \big[\log{(4W/s)} - 2\big].
\end{align}

Upon reducing the gap width $s$, the penetration field diminishes and the
geometrical barrier is more strongly suppressed, see \eq
\eqref{eq:geometric-barrier}.  Vortices become energetically favorable (deep)
inside the sample beyond the equilibrium field (we use \eq \eqref{eq:def-heq}
in the regime $s \ll W$)
\begin{align}\label{eq:heq-double-thin}
   \heq = H_{c1} \frac{d}{2W} \Big\{1 
   - \frac{\log[4\log(4W/s)]+1}{2 \log(4W/s)}\Big\}^{-1},
\end{align}
resulting in a geometric barrier \eqref{eq:eq-geometric-barrier} at $\heq$ which
decreases with $s$,
\begin{align}\label{eq:eq-barrier-double-thin}
   \frac{U_{b}^{\mathrm{eq}}(s)}{\varepsilon_{l} d}= 1 - \frac{\sqrt{d/s}}
              {2 \log(4W/s) - \log[4\log(4W/s)]-1}.
\end{align}
\subsubsection{Penetrated state}\label{sec:sec:sec:Shubnikov-2-tsl}
Increasing the external field beyond its critical value, $\hp$, vortices
penetrate the superconductor from the inner edges at $x = \pm s$ and
accumulate near the position $\a$ inside the strips where the potential
$U_{\mathrm{geo}}(x)$ is minimal. The field and currents take the general form
given in \eqs \eqref{eq:double-field-general} and
\eqref{eq:double-current-general}, with the non-trivial vortex state
determined by the two boundaries of the vortex dome $\bone$ and $\btwo$.
\begin{figure}[t]
\includegraphics[width=0.48\textwidth]{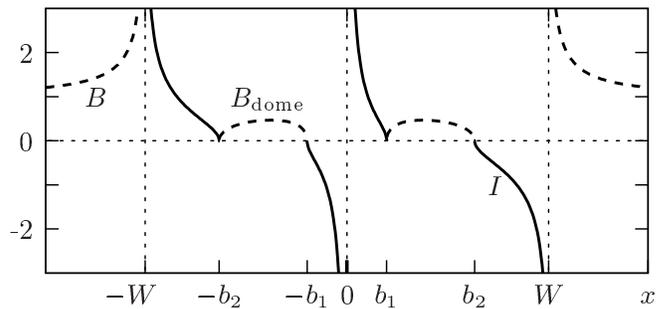}
\caption{Dimensionless field $B_{z}(x)/H$ (dashed line) and current $2\pi
I(x)/H c$ (solid line) as a function of $x$ (for $z=0$) for the same geometry
($w/s = 100$ and $s/d = 100$) as in \fig
\ref{fig:double-strip-meissner-d-ll-s} and an external field $H = 1.2 \hp$
above first penetration. In the penetrated state above $\hp$, vortices
accumulate in a finite region inside each strip (the vortex dome), with
boundaries given by $\pm\bone$ and $\pm\btwo$.}
\label{fig:double-strip-penetrated-d-ll-s}
\end{figure}%
These two parameters satisfy the constraint of vanishing net current in each
strip
\begin{align}\label{eq:double-no-net-current}
   \int_{s}^{\bone} dx\, I(x) + \int_{\btwo}^{W} dx\, I(x) &= 0,
\end{align}
together with the condition [from \eq \eqref{eq:condition-penetration-field}]
\begin{align}\label{eq:critical-condition-penetrated-double}
   \hedge = B_{z}(s-d/2) &= \hc.
\end{align}
While this constraint locks the
field strength at the inner edge to $\hc$, the field strength near
the outer edge continuously grows, but remains below $\hc$. In \fig \ref{fig:double-strip-penetrated-d-ll-s} we show the field and current
profiles in the penetrated state for $H = 1.2\hp$ as obtained from solving
\eqs \eqref{eq:double-no-net-current} and
\eqref{eq:critical-condition-penetrated-double} numerically. The evolution of the dome's boundaries and its width $\btwo - \bone$ with
increasing field is shown in \fig \ref{fig:b1-n-b2}. The maximal field value
in the dome can be estimated with the interpolation formula $B_\mathrm{dome}
\sim H (b_2-b_1)/W$.

In order to find analytic results describing the penetrated state, we have to
simplify the problem of determining the parameters $\bone$, $\btwo$.
Evaluating the condition \eqref{eq:critical-condition-penetrated-double} for
the field \eqref{eq:double-field-general} and expressing the result through
the penetration field $\hp$, the dome boundaries $\bone(H)$ and $\btwo(H)$ are
related via
\begin{align}\label{eq:critical-current-condition}
   \frac{\sqrt{\bone^{2}-s^{2}}\ \btwo}{\a^{2}} = \frac{\hp}{H},
\end{align}
where $\a$ is the (field-independent) zero-current location in the Meissner
state, \eq \eqref{eq:double-zero-current-result}.

It turns out that a perturbative calculation around the penetration field with
the small parameter $h = (H-\hp)/\hp \ll 1$ produces results with a very
limited range of validity. This is due to the rapid growth of the dome width
$\btwo - \bone$ with increasing $h$, leading to a fast break-down of the
approximation.  Approaching the problem from the high field limit $H \gg \hp$
is more successful: starting from the regime where the dome extends over a
large fraction of the strip $\bone \ll \btwo$, we can adopt another
perturbative approach which provides accurate results all the way down to
$\hp$. We use the Ansatz $\btwo = W (1-\nu)^{1/2}$ with $\nu(H) < 1$. For
$s\ll \bone$, where \eq \eqref{eq:critical-current-condition} simplifies to
\begin{align}\label{eq:appr-critical-current-condition}
   \bone = W \frac{\a^{2}}{W^{2}} \frac{\hp}{H} \frac{1}{\sqrt{1-\nu}},
\end{align}
the constraint \eqref{eq:double-no-net-current} of vanishing net current in
the strips can be written as
\begin{align}\label{eq:appr-neutral-current-condition}
   \frac{\hp}{H}\frac{\a^{2}}{W^{2}}
    \left[\log{\Big(\frac{4W}{s}
                    \frac{\a^{2}}{W^{2}}
                    \frac{\hp}{H}
                    \frac{1}{\sqrt{1-\nu}}\Big)}-1\right]&\\[.5em]
   = \E(\sqrt{\nu})-&(1-\nu)\K(\sqrt{\nu}).\nonumber
\end{align}
Solving this equation to leading order in $\hp/H$ where $\nu \ll 1$,
we find that
\begin{align}\label{eq:mu-t-s-l}
   \nu(H) &= \frac{4}{\pi}\frac{\hp}{H}\frac{\a^{2}}{W^{2}}
    \left[\log{\Big(\frac{4W}{s}
                    \frac{\a^{2}}{W^{2}}
                    \frac{\hp}{H}\Big)}-1\right].
\end{align}
\begin{figure}[t]
\includegraphics[width=.48\textwidth]{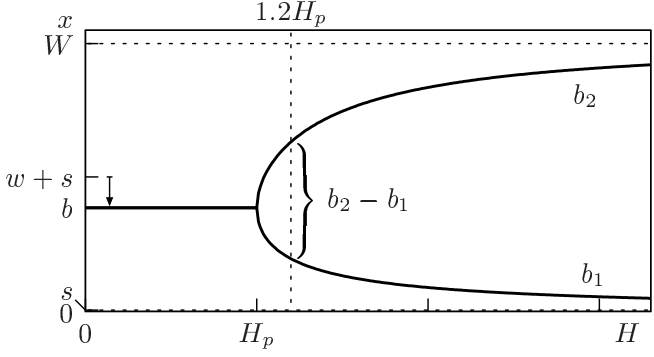}
\caption{Evolution of the dome edges $\bone(H)$ and $\btwo(H)$ with increasing
field $H$ for parameters $w/s = 100$ and $s/d = 100$. For small fields $H
< \hp$, the double strip is in the Meissner phase and $\bone = \btwo = \a$,
where $\a$ is shifted away from the sample center $w+s$.  In the penetrated
state $H>\hp$, vortices accumulate inside the strip and the dome width $\btwo
- \bone$ widens. The field and current profiles for the field $H = 1.2 \hp$
are shown in \fig \ref{fig:double-strip-penetrated-d-ll-s}.}
\label{fig:b1-n-b2}
\end{figure}%
To leading order in $\nu(H)$, the magnetic response in \eq
\eqref{eq:magnetization-via-B-double} takes the form $M \simeq -H \nu(H)
W^{2}/4$, resulting in a logarithmic field-dependence
\begin{align}\label{eq:appr-mag-t-s-l-original-parameters}
   M(H) &\simeq -\frac{\hc}{4\pi} (2W d)
    \bigg\{2\sqrt{\frac{s}{d}}
      \Big[\log{\Big(\frac{4 \hc}{H}\sqrt{\frac{d}{s}\,}\Big)}-1\Big]
    \bigg\}.
\end{align}
\begin{figure}[b]
\includegraphics[width=.43\textwidth]{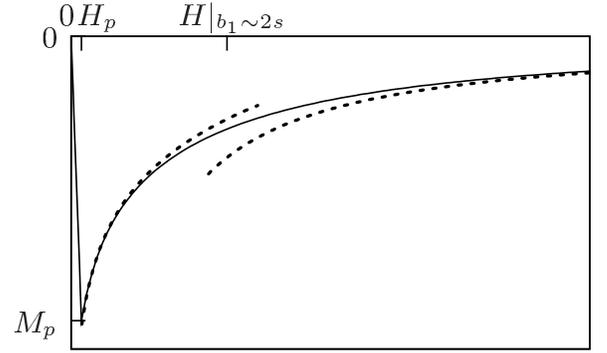}
\caption{Magnetization of a double strip system obtained from numerical
evaluation (solid line) and from the analytic solutions (dashed lines) for
parameters $w/s = 100$ and $s/d = 100$. The expression in \eq
\eqref{eq:appr-mag-t-s-l-original-parameters} is applicable in the field range
$\hp \ll H \ll H|_{\bone\sim2s}$. It turns out, that the analytic
approximation is accurate almost down to $\hp$, where the magnetization is
$M_{p} = M(\hp)$, see \eq \eqref{eq:max-magnetization-double-thin-strip}. For
very large fields, $H > H|_{\bone\sim2s}$, where the distance between the dome
boundary $\bone$ and the sample edge $s$ falls below $s$, the magnetic
response is well described by the asymptotic result in \eq
\eqref{eq:magnetization-in-large-field-assymptotic}. The dome reaches the
edges at a distance $d$ only when $H \sim \hc \gg H|_{\bone \sim 2s}$. Both
approximations \eqref{eq:appr-mag-t-s-l-original-parameters} and
\eqref{eq:magnetization-in-large-field-assymptotic} are shown in their
domain of applicability.}
\label{fig:numeric-vs-analytic}
\end{figure}%
%
Because of the simplification in \eq
\eqref{eq:appr-critical-current-condition}, the validity of the result
\eqref{eq:appr-mag-t-s-l-original-parameters} is limited to fields $H \ll
H|_{\bone\sim2s}\sim (\a^{2}/sW)\hp \approx \sqrt{d/s\,} \hc$, where the
restriction $s \ll \bone$ is satisfied. As shown in \fig
\ref{fig:numeric-vs-analytic}, the expression
\eqref{eq:appr-mag-t-s-l-original-parameters} is in good agreement with the
numerical solution and describes the evolution of the magnetic response over a
large range of fields $\hp \lesssim H \ll H|_{\bone\sim2s}$.

For $\bone - s \ll s$, the same Ansatz $\btwo = W (1-\nu)^{1/2}$ allows to
simplify the constraint \eq \eqref{eq:double-no-net-current} to
\begin{align}\label{eq:no-net-current-for-b1minuss-ll-s}
   \bone &= s + W\nu/2,
\end{align}
while \eq \eqref{eq:critical-current-condition} takes the form
\begin{align}\label{eq:appr-critical-current-condition-2}
   s + W\nu/2 &= \sqrt{s^{2} +
                   W^{2}\Big(\frac{\hp}{H}\frac{\a^{2}}{W^{2}}\Big)^{2}}.
\end{align}
To leading order in $\hp/H$ we find
\begin{align}
   \nu(H) &= \frac{W}{s}\Big(\frac{\hp}{H} \frac{\a^{2}}{W^{2}}\Big)^{2}
\end{align}
and the magnetization reads
\begin{align}\label{eq:magnetization-in-large-field-assymptotic}
   M(H) = -\frac{\hc^{2}}{4H} Wd.
\end{align}
For the strongly penetrated double strip, when the current-carrying regions
are smaller than $s$ but still wider than $d$, the mutual influence of the two
strips becomes negligible. The magnetization
\eqref{eq:magnetization-in-large-field-assymptotic} thus approaches that of
two independent single strips of width $W$ each [see \eq
\eqref{eq:single-penetrated-magnetization}].

Pushing the above `thin strip' solution obtained for $s \gg d$ to the limit $s
= d$, we find for the penetration field in \eq
\eqref{eq:double-penetration-field}
\begin{align}\label{eq:penetration-field-s-is-d-thin-strips}
   \hp &\approx \hc \frac{d}{W} \log{(4W/d)},
\end{align}
which is substantially smaller than that of an isolated strip as given in \eq
\eqref{eq:single-penetration-field}. Similarly, in this limit the
magnetization as approximated by \eq
\eqref{eq:appr-mag-t-s-l-original-parameters} becomes
\begin{align}\label{eq:appr-mag-s-eq-d}
   M(H) &= -\frac{\hc}{4\pi} 4Wd \,\big[\log{(4 \hc/H)}-1\big].
\end{align}
This expression is valid for fields up to $H|_{\bone \sim 2d} \sim \hc$ where
the dome reaches the edges and is consistent with the limit $s 
{\scriptstyle{\ \nearrow\ }} d$ approaching the thickness $d$ from below as
discussed in Sec.\ \ref{sec:finite-thickness} below.

In order to understand the penetration mechanism in the double strip for the
full range of strip separations $2s$, below we extend our analysis to a system
where the gap width $2s$ is much smaller than the thickness $d$ of the strips,
$s \ll d$, see \sect \ref{sec:finite-thickness}. Before doing that, we
briefly elaborate on the corrections due to the vortex structure in the dome.

\begin{figure}[t]
\includegraphics[width=.48\textwidth]{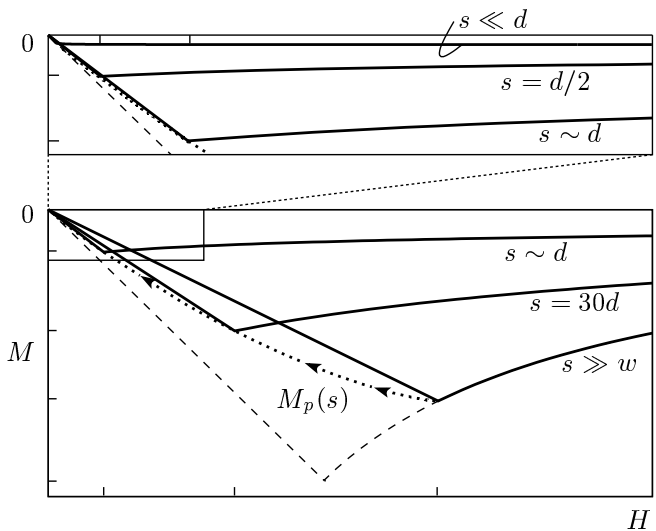}
\caption{The magnetic response of the double strip is shown for different
separation $s$ (solid lines) and a fixed ratio $w/d=10^{3}$. For $s$ larger or
equal to $d$, we adopt the thin-strip approach of \sect
\ref{sec:sec:double-strip}, while for $s < d$ (see top expansion), the
determination of the magnetization curves has to account for the finite
thickness of the strips as described in \sect \ref{sec:finite-thickness}. The
magnetization $M_{p}$ at the penetration field $\hp$ is largest for isolated
strips ($s/w \to \infty$) and reduces upon decreasing $s$. The parametric
curve $(\hp, M_{p})$ as a function of $s$ is indicated by the dotted line. In
the limit $s/w \to 0$ the slope of the magnetization curve in the Meissner
state doubles as compared to that for isolated strips ($s/w \to \infty$).  The
dashed line indicates the magnetization of a single strip of width $2W$,
corresponding to $s = 0$.}
\label{fig:magnetization-s=d-compared-to-single-and-isolated-strip}
\end{figure}%
\subsubsection{Magnetization of the vortex dome}
\label{sec:sec:sec:magnetic-medium-2}
To estimate the quantitative effects arising from the currents around the flux
lines in the vortex dome, we give an upper bound to the corrections of the
magnetic response in \eqs \eqref{eq:appr-mag-t-s-l-original-parameters} and
\eqref{eq:magnetization-in-large-field-assymptotic}. Following the analysis
presented in \sect \ref{sec:sec:sec:magnetic-medium-1}, we find an upper bound
\begin{align}\label{eq:magn-correction-estimated-2}
   \delta M < \frac{H}{4\pi} \frac{H_{c1}}{H_{c1}+H} 2W d
\end{align}
for the magnetization corrections. In the regime $s \ll W$ the
relative correction to the magnetic response is bounded by
\begin{align}\label{eq:relative-corrections-double-1}
   \frac{\delta M}{M} &< \frac{H}{\hc} 
              \frac{H_{c1}}{H_{c1}+H} 
              \sqrt{\frac{d}{4s}}\frac{1}{\log(4\hc/H\sqrt{d/s})-1}
\end{align}
in the low field range $H < \hc \sqrt{d/s}$ [see \eq
\eqref{eq:appr-mag-t-s-l-original-parameters}] and by
\begin{align}\label{eq:relative-corrections-double-2}
   \frac{\delta M}{M} &< \frac{H^{2}}{\hc^{2}}
                         \frac{H_{c1}}{H_{c1}+H} \frac{2}{\pi}
\end{align}
for higher fields field, $H > \hc \sqrt{d/s}$ [see \eq
\eqref{eq:magnetization-in-large-field-assymptotic}]. The first expression
\eqref{eq:relative-corrections-double-1} is always small by the order $d/s$,
while the second expression \eqref{eq:relative-corrections-double-2} predicts
small corrections $\propto (H/\hc)^{2}$ in the field range $H \ll \hc$,
spanning the range of validity for the results presented in this section. We
conclude, that the corrections arising due to the vortex state inside the
superconducting strips are small, justifying the simplified model for the
penetrated state ($\mu = 1$) used in our analysis.

\section{Strips with Finite thickness $\boldsymbol{d}$}
\label{sec:finite-thickness}
\subsection{Introduction}\label{sec:f-t-formalism}
We now explore the double strip geometry for narrow gaps $2s \ll d$. In order
to simplify our discussion, the penetration depth $\lambda$ is assumed to be
negligible\cite{footnote:finite-lambda}, $\lambda \ll s$. With the gap-width $s$ the smallest geometric length and using $d \ll w$, the
results are presented to leading order in $s/d$ and $d/w$, respectively; in
particular, the half-width $W$ of the system is approximated by the width $2w$
of one strip.

The solutions for infinitely thin strips derived in the previous sections have
been regularized near the sample edges with a cut-off $\delta$ of the order of
the thickness, $\delta~\sim~d$. This approach is not appropriate anymore when
the spacial solution near (inside) the gap is determined by the length scale
$s$ rather than $d$. The appropriate boundary conditions then have to be taken
into account on the entire rectangular cross-section and the strips cannot be
treated as infinitely thin anymore.
\begin{figure}[b]
\includegraphics[width=0.45\textwidth]{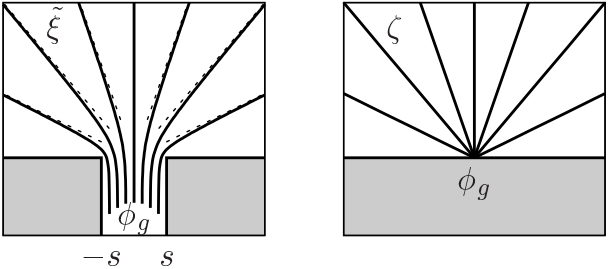}
\caption{Left panel: Field lines for the estuary problem (solid lines in the
$\txi$-plane) as calculated numerically from \eq \eqref{eq:estuary-solution}.
Right panel: the field lines of a point source in the upper half $\zeta$-plane
from which the estuary flow is derived via the inverse Schwarz-Christoffel
transformation $\zeta(\txi)$.  The field lines of the estuary problem approach
that of a point source (dashed lines in the left panel) within a distance $s$
away from the opening.}
\label{fig:estuary-flow}
\end{figure}%

The detailed derivation of the field distribution in the vicinity of the
narrow ($2s$) and elongated ($d$) gap (see \fig \ref{fig:estuary-flow})
presented in \sect \ref{sec:sec:sec:near-field} below will provide us with a
uniform field inside the gap of strength
\begin{align}\label{eq:anticipate-field-in-the-gap}
   B_{g} &= \frac{\phi_{g}}{2s},
\end{align}
where the flux $\phi_{g}$ through the gap has to be determined consistently
with the field distribution far away from the gap. For distances $s <
|\boldsymbol{r}| \ll w$ away from the upper ($+$) and lower ($-$) gap opening,
the field assumes the form of a monopole with radial decay
\begin{align}\label{eq:anticipate-field-away-from-the-opening}
   \boldsymbol{B}(\boldsymbol{r}) &= \pm \frac{\phi_{g}}{\pi} 
   \frac{\boldsymbol{r}}{|\boldsymbol{r}|^{2}}.
\end{align}
The corresponding result expressed through the holomorphic field reads
\begin{align}\label{eq:ant-field-away-from-the-opening}
   \mathcal{B}(\xi) &= i \,\frac{\phi_{g}}{\pi\xi}.
\end{align}
In \sect \ref{sec:sec:sec:far-field} we find the field distribution far away
from the gap, match the far-field solution with the solution in the gap, and
thereby find the flux $\phi_{g}$ through the gap. Along with this derivation,
we will discuss the consequences on the double strip solution originating from
the current and field distribution in and around the gap.

\subsection{Estuary Problem}\label{sec:sec:sec:near-field}
The field distribution inside the gap and near the opening at
$\xi_{{\mathrm{out}}} = 0 + i d/2$ is described by a so-called estuary flow,
i.e., the flow into open space of an incompressible fluid leaving a
canal of width $2s$ and large (infinite) length $d$, see Fig.\
\ref{fig:estuary-flow}. We define the shifted coordinate system $\txi = \xi -
\xi_{{\mathrm{out}}}$ centered at the gap opening two-dimensional estuary
geometry and determine the holomorphic function $\mathcal{B}(\txi)$.
For a diamagnetic superconductor, the field component perpendicular to
the surface vanishes everywhere such that $\mathcal{B}(\txi)$ is purely real
($B_{x} = 0$ and $B_{z} \neq 0$) at the surfaces inside the gap ($\re{[\txi]}
= \pm s$, $\im{[\txi]} <0$) and imaginary ($B_{x} \neq 0$ and $B_{z} = 0$) on
the surfaces along $x$, i.e., for $\im{[\txi]} = 0$ and $|\!\re{[\txi]}|
\geq s$.
\begin{figure}[t]
\includegraphics[width=0.48\textwidth]{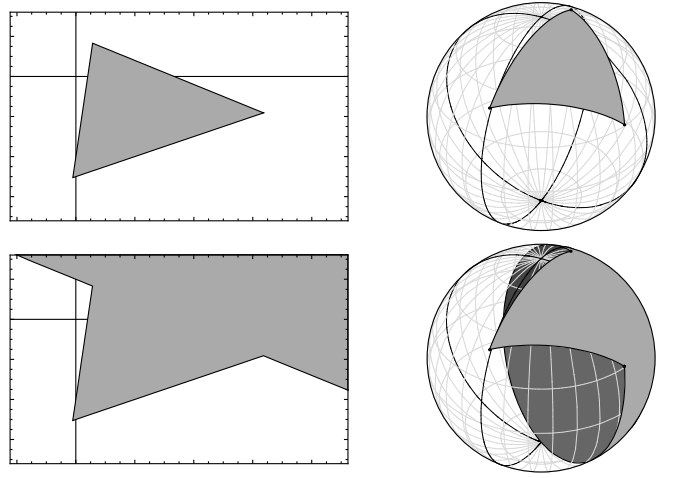}
\caption{Illustration of the stereographic projection of a (conventional)
triangle from the Euclidean plane (top left) to the Riemann sphere (top
right).  Replacing one edge of this triangle by its complement (passing
through infinity) generates an unbound triangle. This situation is
illustrated both in the Euclidean plane (bottom left) and on the Riemann sphere
(bottom right) after a stereographic projection.  }
\label{fig:stereographic-projection-example}
\end{figure}%

\begin{figure}[t]
\includegraphics[width=0.48\textwidth]{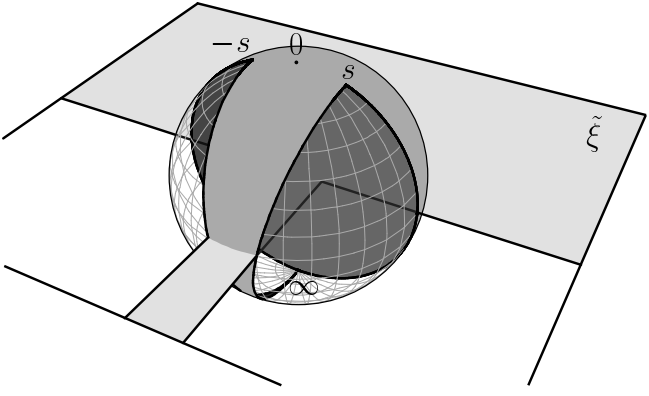}
\caption{Visualization of the field-allowed region (light gray) of the estuary
geometry on the Riemann sphere (filled region) via a stereographic projection.
The triangular shape of the boundary of the estuary, with vertices $\txi_{v}$
at $\pm s$ and $-i\infty$ is clearly visible on the Riemann sphere
representation, see also \fig \ref{fig:stereographic-projection-example}.
Here, the north pole corresponds to the origin of the complex plane $\txi$,
while the complex infinity is projected onto the south pole.  A line in the
original plane $\txi$ is mapped to a circle (passing through the south pole)
on the Riemann sphere.}
\label{fig:stereographic-projection}
\end{figure}

This boundary value problem can be solved with the help of a
Schwarz-Christoffel transformation\cite{Henrici_74} describing a biholomorphic
mapping of the upper complex half-plane $\zeta$, $\im{[\zeta]} \geq 0$, onto
the inner of a polygon. Indeed, the field-allowed region in the estuary
geometry is a special case of an unbounded triangle (visualized in Figs.\
\ref{fig:stereographic-projection-example} and
\ref{fig:stereographic-projection}), with vertices $\txi_{v}$ at $-s$,
$-i\infty$, and $s$ and internal angles $3\pi/2$, $0$, and $3\pi/2$. The
corresponding Schwarz-Christoffel transformation takes the form
\begin{align}\label{eq:sc-transformation}
   \txi(\zeta) &= s + \frac{2}{\pi}
                     \bigg[\sqrt{\zeta^{2}-s^{2}}
                     - 2s \arctan{\sqrt{\frac{\zeta - s}{\zeta + s}}}\bigg]
\end{align}
and maps the upper half-plane $\zeta$ (\fig \ref{fig:estuary-flow}, right) to
the estuary plane $\txi$ (\fig \ref{fig:estuary-flow}, left). The flux
$\phi_{g}$ emanating from the vertex at $\txi_{v} = -i \infty$ in the estuary
is conserved in the transformation \eq \eqref{eq:sc-transformation} and maps
to a point source of strength $\phi_{g}$ at $\zeta = 0$, with field lines
dispersing into the upper half plane $\im[\zeta] \geq 0$. The complex
potential\cite{Thomson_60}
\begin{align}\label{eq:complex-potenital}
   \bar\Omega(\zeta) = \frac{i\phi_g}{\pi} \log\zeta
\end{align}
is generating the field $\mathcal{\bar B}(\zeta)= d\bar\Omega/d\zeta = i\phi_{g}/\pi \zeta$ of this point source in the upper half-plane. Transforming back to the estuary geometry, the potential $\Omega(\txi) = \bar
\Omega[\zeta(\txi)]$ generates the field
\begin{align}\label{eq:estuary-solution}
   \mathcal{B}(\txi)= \frac{d\Omega}{d\txi} = \frac{i\phi_g}{2} \frac{1}{\sqrt{\zeta(\txi)\big.^2-s^2}}.
\end{align}
The last equality was obtained by using the Schwarz-Christoffel transformation
\eqref{eq:sc-transformation}. Alternatively, the analysis on the level of
fields involves the solution $\mathcal{\bar B}(\zeta) = i \phi_{g}/\pi \zeta$
for a point source and the transformation back involves an additional
derivative, $\mathcal{B}(\txi) = (d\zeta/d\txi) \mathcal{\bar
B}[\zeta(\txi)]$. In \fig \ref{fig:estuary-flow}, we show the resulting field
lines of \eq \eqref{eq:estuary-solution}  as obtained from inverting \eq
\eqref{eq:sc-transformation} numerically.

In our further discussion it is sufficient to determine the field distribution
in the asymptotic regimes where analytic results are available. Deep inside
the gap ($-\im{[\txi]} \gg s$) the inverse of \eq
\eqref{eq:sc-transformation} takes the form
\begin{align}\label{eq:gap-inversion}
   \zeta(\txi) &= 2s \, e^{- [i \pi (\txi - 1)/ 2s] -1}
\end{align}
and using \eq \eqref{eq:estuary-solution}, we find [up to corrections $\propto
\exp({\pi \tilde{z}/2s})$] a uniform field directed along $z$ of strength
\begin{align}\label{eq:field-in-the-gap}
   B_{g} &= \frac{\phi_{g}}{2s}.
\end{align}
Near the corner of the estuary, $|\txi - s| \ll s$, the transformation \eq
\eqref{eq:sc-transformation} reads
\begin{align}\label{eq:sc-near-corner-s}
   \frac{\txi-s}{2s} &\sim \frac{2}{3\pi} \left(\frac{\zeta-s}{2s}\right)^{3/2}
\end{align}
and a similar expression is found near $\txi = -s$. For both corners, the
holomorphic field \eqref{eq:estuary-solution} shows a power law singularity
$\propto |\txi \pm s|^{-1/3}$, which will be regularized in a real sample by
the partial penetration (at a depth $\sim s$) of vortices into the sample
corners.

Far away from the opening, $|\txi| \gg s$ and $\im{[\txi]} \geq 0$, the
inverse transform becomes $\zeta(\txi) = \pi \txi / 2$ and the holomorphic
function \eqref{eq:estuary-solution} assumes the limiting form
\begin{align}\label{eq:field-away-from-the-opening}
   \mathcal{B}(\txi) &= i \,\frac{\phi_{g}}{\pi\txi}
\end{align}
describing a point source of strength $\phi_{g}$ located at $\txi = 0$.

\subsection{Narrow gap double strip}\label{sec:sec:ngds}
\label{sec:sec:sec:far-field}
Away from the gap and from the outer strip edges, a thin-strip description
similar to the one discussed in \sect \ref{sec:thin-strips} is applicable, with
the holomorphic field taking the form
\begin{align}\label{eq:double-field-general-narrow-gap}
   \mathcal{B}(\xi) &= H \sqrt{\frac{(\xi^{2}-\bone^{2})(\xi^{2}-\btwo^{2})}
                                    {\xi^{2}(\xi^{2}-W^{2})}}.
\end{align}
The factor $\xi^{2}$ in the denominator [replacing $(\xi^{2}-s^{2})$ in \eq
\eqref{eq:ant-double-strip-function}] captures the flux emanating from the
point-like source as derived in \eq \eqref{eq:field-away-from-the-opening}.
From the above expression, the field distribution along the $x$-axis is given
by
\begin{align}\label{eq:double-field-general-narrow-gap_xz}
   \frac{B_{z}(x)}{H} &= \left\{ \begin{aligned}
            &\sqrt{\frac{(x^{2}-\bone^{2})(\btwo^{2}-x^{2})}
                 {x^{2}(W^{2}-x^{2})}},
             && \mathrm{for\ } \bone \leq x \leq \btwo,\\
            &\sqrt{\frac{(x^{2}-\bone^{2})(x^{2}-\btwo^{2})}
                 {x^{2}(x^{2}-W^{2})}},
             && \mathrm{for\ } W \leq x.
            \end{aligned}\right.
\end{align}
Comparing \eqs \eqref{eq:field-away-from-the-opening} and
\eqref{eq:double-field-general-narrow-gap} in the regime $|\xi| \ll \bone$, we
find the flux
\begin{align}\label{eq:flux-general-narrow-gap}
   \phi_{g}= H W \pi \bone \btwo/W^{2}
\end{align}
and the uniform field strength \eqref{eq:field-in-the-gap} inside the gap $|x|
< s$ takes the form
\begin{align}\label{eq:field-general-in-the-gap}
   B_{g} = H \frac{\pi W}{2 s} \frac{\bone \btwo}{W^{2}}.
\end{align}
Note, that for $s \ll d \ll w$, the difference between the shifted coordinate
$\txi$ and $\xi$ is beyond our resolution, such that $\txi = \xi$.

The current contribution from the region away from the gap is obtained from
the holomorphic field in \eq \eqref{eq:double-field-general-narrow-gap} via
Amp\`eres law \eqref{eq:ampere-law} and reads
\begin{align}\label{eq:double-current-general-narrow-gap}
   I(x) &= \left\{\begin{aligned}
            &\frac{H c}{2\pi}
                 \sqrt{\frac{(\bone^{2}-x^{2})(\btwo^{2}-x^{2})}
                 {x^{2}(W^{2}-x^{2})}},
             && s \leq |x| \leq \bone,\\
            &\frac{-H c}{2\pi}
                 \sqrt{\frac{(x^{2}-\bone^{2})(x^{2}-\btwo^{2})}
                 {x^{2}(W^{2}-x^{2})}},
             && \btwo \leq |x| \leq W,\\
            &0,
             && \mathrm{otherwise}.
            \end{aligned}\right.
\end{align}
The $1/x$ dependence of the current is applicable only for $|x| \gg s$. However,
it turns out that the deviation of $I(x)$ (as obtained from solving \eq
\eqref{eq:sc-transformation} numerically) from $1/x$ is not relevant for the
further analysis, and the expression given above for the sheet current density
$I(x)$ can be used down to $|x| = s$.

The homogeneous field \eqref{eq:field-general-in-the-gap} inside the gap is
generated by a screening current density
\begin{align}\label{eq:current-density-in-the-gap}
   j_{g}(x,|z|<d/2) &= \frac{x}{|x|} \frac{B_{g}\, c}{4\pi} \,\delta(|x|-s)
\end{align}
flowing along $y$ at the gap surfaces ($x=\pm s$, $|z| \leq d/2$).  Here
$\delta$ is the Dirac delta function, which accounts for the assumption
$\lambda \to 0$.  The two current channels at $x = \pm s$ provide a
significant contribution to the total current in the strips. Note that these
channels exist for $s \gg d$ as well; for large gaps their contribution to the
total current is negligible, though. To treat these currents on equal footing
with the sheet current flowing in the strips ($s \leq x \leq W$) we define the
sheet current density for the gap currents
\begin{align}\label{eq:current-in-the-gap}
    I_{g}(x) &= \frac{x}{|x|} 
    \frac{B_{g} c}{4\pi} \,d\,\delta(|x|-s)
\end{align}
by integrating \eq \eqref{eq:current-density-in-the-gap} over the strip
thickness $d$.

\begin{figure}[t]
\includegraphics[width=0.45\textwidth]{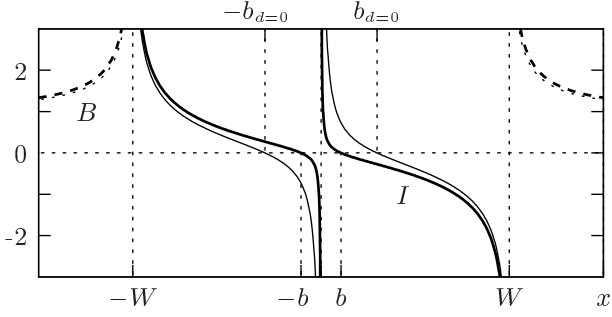}
\caption{Dimensionless current $2\pi I(x)/H c$ (thick solid line) and reduced
magnetic field $B_{z}(x)/H$ (dashed line) as a function of $x$ and for $z=0$
in a double strip in the Meissner state. The profiles are calculated for the
case $s \ll d \ll 2w$ with the parameters $w/d = d/s = 100$. The current
profile inside the strips changes sign at $\pm \a$, with $|\a| \approx 0.21
w$. The additional current $I_{g}(x)$ from \eq \eqref{eq:current-in-the-gap}
flowing near the inner edges of the strips changes the condition of zero net
current \eqref{eq:no-net-current-explicit} dramatically. The thin lines show
the current and field profiles of the double strip in the thin strip limit,
$d/s \to 0$, for fixed $s = 10^{-4}w$.}
\label{fig:double-strip-meissner-s-ll-d}
\end{figure}

The currents flowing along the vertical surfaces at the outer edges ($|x| =
W$) are parametrically smaller as compared to the contributions near the gap
($|x| = s$) and are neglected here. The two dominant current contributions
then add up to the total current, $I_{\mathrm{tot}}(x) = I(x) +
I_{g}(x)$. This current distribution, when compared to the thin
strip case, corresponds to a rearrangement of the current densities towards
the inner edges of the strips, see \fig
\ref{fig:double-strip-meissner-s-ll-d}.

The diamagnetic contribution from the current $I(x)$ in the strip,
\begin{align}\label{eq:double-magnetization-general-narrow-gap}
   M(H) &= -\frac{H}{4}(W^{2}-\bone^{2}-\btwo^{2}),
\end{align}
as obtained from evaluating \eq \eqref{eq:magnetization-through-B} with the
field \eqref{eq:double-field-general-narrow-gap}, is parametrically larger
($\propto W/d$) than the paramagnetic contribution
\begin{align}\label{eq:magnetization-in-the-gap}
   M_{g}(H) &= \frac{H}{4} W^{2} \frac{\bone \btwo d}{W^{3}}
\end{align}
from the current $I_{g}(x)$ along the gap surface and we neglect the latter in
the following.
The problem is then, once again, reduced to finding the parameters $\bone$ and
$\btwo$ within the Meissner- and penetrated states.
\subsubsection{Meissner state}\label{sec:sec:sec:Meissner-3}
\begin{figure}[t]
\includegraphics[width=0.45\textwidth]{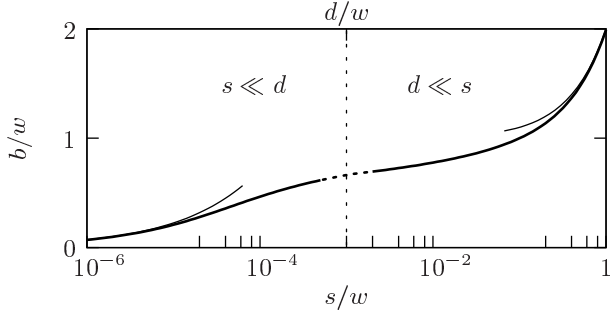}
\caption{The parameter $\a$ characterizing the Meissner phase of the double
strip is plotted against the half-width $s$ of the gap between the strips.
All lengths are normalized to the half width $w$ of the strips.  The fixed
strip thickness $d = 10^{-3}w$ separates two regimes; in the thin strip regime
$d \ll s$, $\a$ depends on $s$ via \eq \eqref{eq:double-zero-current-result}.
For $s \ll d$, $\a(s)$ is given through \eq \eqref{eq:a-over-W-ratio-2}. In
between, i.e., for $s \sim d$, a smooth cross-over (dashed line)
connects the two limits.  In both far asymptotic limits $s \lll d$ and $s \ggg
d$, the position $\a(s)$ follows a simple behavior $\a(s) = 2w\sqrt{4s/\pi d}$
and $\a(s) = w + s$, respectively (thin lines).}
\label{fig:a-of-s}
\end{figure}

In the Meissner state were $\bone = \btwo = \a$, the field $B_{z}$ [\eq
\eqref{eq:double-field-general-narrow-gap}] along the $x$-axis simplifies to
\begin{align}\label{eq:double-field-meissner-narrow-gap}
   B_{z}(x) &=
    H \frac{x^{2}-\a^{2}}
                 {|x| \sqrt{x^{2}-W^{2}}}.
\end{align}
for $|x|>W$ and is constant [\eq \eqref{eq:field-general-in-the-gap}],
\begin{align}\label{eq:field-meissner-in-the-gap}
   B_{g} = H \pi \frac{\a^{2}}{2 s W},
\end{align}
inside the gap ($|x|<s$). The total sheet current density reads
\begin{align}\label{eq:double-current-meissner-narrow-gap}
   I_{\mathrm{tot}}(x) &= - \frac{H c}{2\pi}\bigg[
                         \frac{x^{2}-\a^{2}}{x \sqrt{W^{2}-x^{2}}}
                       - \frac{x}{|x|}
                         \frac{\pi \a^{2}}{4sW} \,d\,\delta(|x|-s) \bigg]
\end{align}
and the general expression \eqref{eq:double-magnetization-general-narrow-gap}
for the magnetization takes the form
\begin{align}\label{eq:double-magnetization-meissner-narrow-gap}
   M(H) &= -\frac{H}{4}(W^{2}-2\a^{2}).
\end{align}
The value of the parameter $\a$ is fixed by the constraint of vanishing net
current given as
\begin{align}\label{eq:no-net-current-explicit}
   \int\limits_{s}^{W} \frac{dx \, x}{\sqrt{W^{2}-x^{2}}}
        = \a^{2} \bigg[ \frac{\pi d}{4sW} + \int\limits_{s}^{W} 
                 \frac{dx}{x \sqrt{W^{2}-x^{2}}}\bigg].
\end{align}
The above integrals simplify in the limit $s \ll W$ and the parameter $\a$
takes the asymptotic form
\begin{align}\label{eq:a-over-W-ratio-2}
   \a^{2} &= \frac{W^{2}}{\pi d/4 s + \log\big(2W/s\big)}.
\end{align}
In contrast to the result for thin strips [see \eq \eqref{eq:a-over-W-ratio}],
where $\a^{2}$ changes logarithmically with $s$, in the present case the
dependence on $s$ is dominated by the linear term $d/s$ in the denominator. As
a result,  the parameter $\a$ is substantially reduced when $s \ll d$, see
\fig \ref{fig:a-of-s}, which is due to the additional currents $I_g$ flowing
at the (vertical) gap surface and producing a substantial rearrangement of the
overall current density as shown in \fig
\ref{fig:double-strip-meissner-s-ll-d}.

We note that the numerical factor $\pi/4$ of the term $d/s$ in the above
expression is precisely known since it derives from the current $I_{g}(x)$
originating from screening the uniform field inside the gap, \eq
\eqref{eq:current-in-the-gap}. The prefactor under the logarithm, however,
will be modified if the field distribution at the opening of the estuary is
accurately taken into account. Indeed, approaching the corner $(s,d/2)$ from
both surfaces $(x,d/2)$ and $(s,z)$ the field deviates from the assumed
behavior $B_{x}(x)\propto 1/x$ and $B_{z}(z) = \mathrm{const}$ [following from
\eqs \eqref{eq:double-field-general-narrow-gap} and
\eqref{eq:field-general-in-the-gap} respectively]. The precise field
distribution (and its related current profile) can be derived by solving \eq
\eqref{eq:sc-transformation} numerically and inserting the result into \eq
\eqref{eq:estuary-solution}. Neglecting partial penetration of the edge
corners, \eq \eqref{eq:a-over-W-ratio-2} will be modified to
\begin{align}
   \a^{2} &= \frac{W^{2}}{\pi d/4 s + \log\big(2.38 W/s\big)}.
\end{align}
Since the precision of this expression also suffers from corrections (e.g.
from partial penetration of the edge corners), we will use the relation
\eqref{eq:a-over-W-ratio-2} in the following.

The diamagnetic response in the Meissner phase follows from
\eqref{eq:double-magnetization-meissner-narrow-gap} and reduces to
\begin{align}\label{eq:mag-meissner-double-strip-s-ll-d}
   M(H) &\approx -\frac{H}{4}W^{2} \bigg[1-\frac{8s/\pi d}{1 + (4s/\pi
   d)\log\big(2W/s\big)}\bigg].
\end{align}
This result approaches that of a single strip of double width [\cf \eq
\eqref{eq:single-magnetization-meissner} with $w \to W$] upon reducing $s$ far
below $d$.

Using \eq \eqref{eq:a-over-W-ratio-2} in the expression
\eqref{eq:flux-general-narrow-gap} for the flux through the gap, we find that
\begin{align}\label{eq:flux-meissner-finite-thickness}
   \phi_{g} &= H 2W \frac{2s/d}{1 + (4s/\pi d)\log\big(2W/s\big)}
\end{align}
shrinks (up to logarithmic corrections) linearly with decreasing $s$ (note
that $\bone \btwo = \a^{2}$). Compared to the blocked flux $\phi_{b} \approx
H\, 2W$, only a small fraction $\sim 2s/d$ passes through the narrow gap of
width $2s$ and length $d$. Consequently, the field strength inside the gap,
\begin{align}\label{eq:double-field-meissner-in-the-gap}
   B_{g} = H \frac{2W}{d} \frac{1}{1 + (4s/\pi d)\log\big(2W/s\big)},
\end{align}
does not diverge for $s/d \to 0$ but saturates at $H\, 2W/d$.

As the field profile inside the gap from where vortices start penetrating the
sample is precisely known, the present penetration process is more accurately
described than the one for the thin-strip limit where the strips are separated
by a distance larger than $d$. Corrections originating from the precise field
distribution near the opening of the estuary affect the results only to the
next-to-leading order.

As before, the penetration starts when the field inside the gap reaches the
strength $\hc$, i.e., at the penetration field
\begin{align}\label{eq:double-penetration-field-narrow-gap}
   \hp &= \hc \frac{2s}{\pi W} \frac{W^{2}}{\a^{2}}
        = \hc \frac{d}{2W}
		  \Big[1 + \frac{4s}{\pi d} \log \Big(\frac{2W}{s}\Big)\Big].
\end{align}
In the limit $s/d \to 0$, the penetration field asymptotically reaches the
value $\hc d/2W$, that is the penetration field of the elliptic strip, \cf \eq
\eqref{eq:single-penetration-field-ellipse}. The retardation of field
penetration originating from the geometrical barrier has completely
disappeared in this limit. At penetration, $H = \hp$, the diamagnetic
response
\begin{align}\label{eq:magnetization-at-penetration-point}
   M_{p} &= -\frac{\hc}{16}(2Wd) 
   \Big\{1 + \frac{2s}{\pi d} \Big[2\log \Big(\frac{2W}{s}\Big)-1\Big]\Big\}\\
   &= -\frac{\hp}{4}W^{2} \Big[1-\frac{8s/\pi d}{1 + (4s/\pi
   d)\log\big(2W/s\big)}\Big],
\end{align}
has collapsed by a factor $\sim (d/W)^{1/2}$ as compared to that of a single
strip, \eq \eqref{eq:single-m-at-penetration-field}, for which the geometrical
barrier is fully active. This so-called `suppression of the geometrical
barrier', the collapse of $\hp$ and of $M(H)$, is a central result of this
work.  Although in the limiting case $s/d \to 0$, the Meissner response and
the field of first penetration coincide with that of an elliptically shaped
strip, beyond $\hp$, the magnetic signatures of the double strip still differs
substantially from those of the elliptic sample, see Figs.\
\ref{fig:numeric-vs-analytic-f-t-l} and \ref{fig:descending-double-strip} as
well as the discussion below.

Upon decreasing the gap width $s$, the penetration field $\hp$ is reduced,
what leads to a stronger suppression of the geometrical barrier as follows
from \eq \eqref{eq:geometric-barrier}. The calculation of the equilibrium
field defined through \eq \eqref{eq:def-heq} provides the result
\begin{align}\label{eq:double-eq-field-narrow-gap}
   \heq = H_{c1} \frac{d}{2W} \Big[1 - \frac{\log(W^{2}/b^{2})-1}
                {\pi d / 2s + \log(4W^{2}/s^{2})}\Big]^{-1},
\end{align}
approaching $H_{c1} d/2W$ and the corresponding geometrical barrier height \eqref{eq:eq-geometric-barrier} vanishes as $s \log(s)$,
\begin{align}
   U_{b}^{\mathrm{eq}} &= \varepsilon_{l} d \Big(1 -
   \Big\{1 + \frac{2s}{\pi d}
                   \big[\log(4\a^{2}/s^{2})-1\big]\Big\}^{-1}\Big)\\
   &\approx \varepsilon_{l} d\ \frac{2s}{\pi d}
                   \big[\log(16W^{2}/\pi s d)-1\big],
\end{align}
where we have assumed that $s \log(W/s) \ll d$ for the last equality.

\subsubsection{Penetrated state}\label{sec:sec:sec:Shubnikov-2-ft}
The field and current distributions \eqs
\eqref{eq:double-field-general-narrow-gap},
\eqref{eq:field-general-in-the-gap} and \eqs
\eqref{eq:double-current-general-narrow-gap}, \eqref{eq:current-in-the-gap}
describe the penetrated state once the parameters $\bone$ and $\btwo$ have
been found; the latter have to respect the limits $\bone-s \gg s$ and
$W-\btwo \gg d$ and are determined by the usual conditions governing the
evolution of the vortex dome, the vanishing of the total currents in the
strips,
\begin{align}\label{eq:no-net-current-2}
   \int\limits_{s_{-}}^{\bone} dx\, I_{\mathrm{tot}}(x)
      + \int\limits_{\btwo}^{W} dx\, I_{\mathrm{tot}}(x) &= 0
\end{align}
and the condition of criticality at the edge regulating the vortex entrance,
here $B_{g} = \hc$. The latter condition is equivalent to the requirement that
the flux $\phi_{g}$ in \eqref{eq:flux-general-narrow-gap} saturates at $\hc
2s$, or
\begin{align}\label{eq:critical-field-condition}
   \frac{\bone \btwo}{\a^{2}} &= \frac{\hp}{H},
\end{align}
as expressed through the penetration field $\hp$ and the zero-current position
$b$ of the Meissner state.

\begin{figure}[t]
\includegraphics[width=0.45\textwidth]{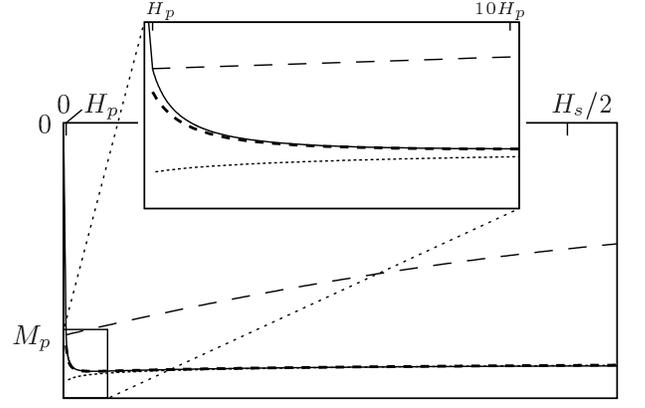}
\caption{The diamagnetic response $M(H)$ (solid line) for the double strip in
the limit $s \ll d \ll w$ (here $w/d = d/s = 100$) as obtained from solving
\eqs \eqref{eq:no-net-current-2} and \eqref{eq:critical-field-condition}
numerically. In addition the dotted (dashed) line shows the analytic result
for the magnetization $M(H) = -H \nu(H) W^{2}/4$, where $\nu(H)$ is obtained
from solving \eq \eqref{eq:appr-neutral-current-condition-f-t-l} to linear
(quadratic) order in $\hp/H$, see \eq \eqref{eq:mu-second-order-f-t-l-0}. It
is necessary to express the solution to second order in $\hp/H$ as the first
order solution gives only poor results close to $\hp$. The magnetization of a
single elliptic strip of width $2W$ and thickness $d$ (long thin dashes) is
reversible and with linear slope beyond $\hp$ (as also shown in \fig
\ref{fig:elliptic-magnetization}).}
\label{fig:numeric-vs-analytic-f-t-l}
\end{figure}

As before, the perturbative calculation around the penetration field $\hp$ is
very limited due to the rapid growth of the dome width $\btwo-\bone$ beyond
$\hp$ and we concentrate on the high-field expansion where the vortex domes
occupy a large fraction of the strips $\bone \ll \btwo$, providing results
over a large field-range. The two conditions regulating the dome evolution
then can be simplified and an analytic solution can be given. With the Ansatz
$\btwo = W (1-\nu)^{1/2}$ with $\nu < 1$, the inner dome edge
\begin{align}\label{eq:bone}
   \bone &= \frac{\a^{2}}{W} \frac{\hp}{H} \frac{1}{\sqrt{1-\nu}}
\end{align}
is expressed through $\nu$ with the help of \eq
\eqref{eq:critical-field-condition}. Assuming $s \ll \bone$ and $\bone \ll
\btwo$, the requirement of vanishing net current in \eq
\eqref{eq:no-net-current-2} simplifies to
\begin{align}\label{eq:appr-neutral-current-condition-f-t-l}
   \frac{\hp}{H}\frac{\a^{2}}{W^{2}}
    \left[\frac{W^{2}}{\a^{2}} +
          \log{\Big(\frac{\a^{2}}{W^{2}}
                    \frac{\hp}{H}
                    \frac{1}{\sqrt{1-\nu}}
               \Big)}-1\right] &\\[.5em]
   = \E(\sqrt{\nu})-(1&-\nu)\K(\sqrt{\nu}).\nonumber
\end{align}
For large fields $H \gg \hp$, where $\nu$ is small, the above equation can be
expanded in $\nu$. Solving for $\nu(H)$ to second order in $\hp/H$, we obtain
\begin{align}\label{eq:mu-second-order-f-t-l-0}
  \nu(H) &\approx \frac{4}{\pi}\frac{\hp}{H}
   \bigg\{ 1 + \frac{\a^2}{W^2}
   \bigg[\log{\left(\frac{\a^2}{W^2}\frac{\hp}{H}\right)}-1\bigg]
   \bigg\}\\
   &\ - \frac{2}{\pi^2}\frac{\hp^2}{H^2}
       \bigg\{1+\frac{\a^2}{W^2}
       \bigg[\log{\bigg(\frac{\a^2}{W^2}\frac{\hp}{H}\bigg)}-1\bigg]
       \bigg\}^2\nonumber\\
   &\ + \frac{8}{\pi^2}\frac{\hp^2}{H^2}
     \frac{\a^2}{W^2}
       \bigg\{1+\frac{\a^2}{W^2}
         \bigg[\log{\bigg(\frac{\a^2}{W^2}\frac{\hp}{H}\bigg)}-1\bigg]
       \bigg\}.\nonumber
\end{align}
The magnetic response given in \eq
\eqref{eq:double-magnetization-general-narrow-gap} simplifies to $M(H) = - H
\nu(H)W^{2}/4$ and \fig\ref{fig:numeric-vs-analytic-f-t-l} shows the result of
combining this expression with $\nu(H)$ from \eq
\eqref{eq:mu-second-order-f-t-l-0}.  Although the range of applicability $\hp
\ll H$ of the above expression does not a-priori cover the regime near
penetration, the results are still in good agreement with the numerical
solution down to $H \approx \hp$.

Neglecting irrelevant terms of order $(\a/W)^{2}(\hp/H)^{2}$ and higher in \eq
\eqref{eq:mu-second-order-f-t-l-0}, the magnetization reads
\begin{align}\label{eq:mag-second-order-f-t-l}
   M(H) &\approx  \bar{M}\bigg\{
   1 + \frac{\a^2}{W^2} \bigg[
                           \log{\bigg(\frac{\a^2}{W^2}\frac{\hp}{H} \bigg)}-1
                        \bigg]
    - \frac{1}{2\pi}\frac{\hp}{H}
                        \bigg\},
\end{align}
with $\bar{M} = -\hp W^{2}/\pi$. In contrast to the thin strip case [see \eq
\eqref{eq:appr-mag-t-s-l-original-parameters}], where the magnetization
depends logarithmically $\propto \log{(\hc/H)}$ on the applied field, the
magnetic response in the present limit is dominated by a field-independent
contribution,
\begin{align}\label{eq:appr-mag-f-t-l}
   \bar{M} &\approx -\frac{\hp}{\pi} W^{2} \approx -\frac{\hc}{4\pi} (2Wd),
\end{align}
producing an almost flat magnetization. This flatness is the result of the
particular current distribution inside the strips: The current flowing close
to the inner edge is dominated by the contribution $I_{g}(x)$ from the gap,
i.e.,
\begin{align}\label{eq:current-estimation-near-the-gap}
   \int\limits_{s_{-}}^{\bone} dx\, I_{\mathrm{tot}}(x) 
   &\approx \int\limits_{s_{-}}^{s_{+}} dx\, I_{g}(x) 
   = \frac{\hc c}{4\pi} \,d.
\end{align}
To satisfy the condition \eqref{eq:no-net-current-2} of vansihing net current,
the current density $I(x)$ between $\btwo$ and $W$ has to compensate the gap
contribution, leading to
\begin{align}\label{eq:current-estimation-outer-edge}
   \int\limits_{\btwo}^{W} dx\, I_{\mathrm{tot}}(x) 
   &\approx -\frac{\hc c}{4\pi} \,d.
\end{align}
Once the dome occupies a large fraction of the sample, these currents flow at
the outer edge, i.e., a distance $\sim W$ away from the origin and produce the
dominant (field-independent) contribution $-\hc(2Wd)/4\pi$ [\cf \eq
\eqref{eq:appr-mag-f-t-l}] to the magnetization at large fields (the factor 2
originates from the integration over both strips). Note that in the limit $s/d
\to 0$, the leveling out of the magnetization at the value given in \eq
\eqref{eq:appr-mag-f-t-l} is by a factor $4/\pi$ larger than its value at
penetration $\hp$ [see \eq \eqref{eq:magnetization-at-penetration-point}].

Although almost constant, the magnetization \eqref{eq:mag-second-order-f-t-l}
assumes a maximal diamagnetic signal
\begin{align}\label{eq:max-magnetization}
   M(H_{m}) &= \bar{M} \bigg\{1 + \frac{4s}{\pi d} 
   \bigg[\log\bigg(\frac{32s^{2}}{\pi d^{2}}\bigg) -2\bigg]\bigg\}
\end{align}
at the applied field
\begin{align}\label{eq:H-max}
   H_{m} &\approx \hp \frac{d}{8s}
          \approx  \frac{\hc}{16}\frac{d^{2}}{s W}.
\end{align}
For $s/d \lesssim d/16W$, the diamagnetic response monotonically increases up
to $H \sim \hc$. On the other hand, we may extrapolate the expression
\eqref{eq:H-max} to $s \lesssim d$ and predict a value of the gap parameter $s
\sim d/8$ where $H_{m}$ merges with the penetration field $\hp$ upon
increasing $s$. The ``flatness'' of the magnetization curve in the penetrated
state is quantified by relating the slope $M'(H)$ [as obtained from \eq \eqref{eq:mag-second-order-f-t-l}] to the Meissner slope $-
W^{2}/4$, yielding
\begin{align}\label{eq:relative-magnetization-slope}
   -\frac{4 M'(H)}{W^{2}} = \frac{2}{\pi^{2}} 
    \bigg(\frac{\hp^{2}}{H^{2}} - \frac{8s}{d}\frac{\hp}{H}\bigg) \ll 1.
\end{align}

As for thin strips and wide gaps (see \sect
\ref{sec:sec:sec:Shubnikov-2-tsl}), we can push the results obtained in the
limit $s \ll d$ to the extreme case $s \to d$. The penetration field
\begin{align}\label{eq:penetration-field-s-is-d-thick-strips}
   \hp &\approx \hc \frac{d}{2W}
          \Big[1 + \frac{4}{\pi} \log \Big(\frac{2W}{d}\Big)\Big]
\end{align}
as obtained from \eq \eqref{eq:double-penetration-field-narrow-gap} with $s = d$, agrees up to numbers of order unity with the result obtained from the opposite
limit $s \gg d$, see \eq
\eqref{eq:penetration-field-s-is-d-thin-strips}. Taking the limit $s \to d$ from the regime of small gaps $s \ll d$, the magnetization is dominated by the first term in \eq
\eqref{eq:mu-second-order-f-t-l-0} and simplifies to
\begin{align}\label{eq:magnetization-s-is-d-thick-strips}
   M(H) &\approx -\frac{\hc}{4\pi} (4Wd) \Big[\frac{2}{\pi}
   \log{\Big(\frac{4\hc}{\pi H}\Big)} + \frac{4 - \pi}{2\pi} \Big].
\end{align}
This expression agrees well with the corresponding expression
\eqref{eq:appr-mag-s-eq-d} obtained in the limit $s {\scriptstyle{\ \searrow\
}}d$ approaching the thickness $d$ from above.

\begin{figure}[t]
\includegraphics[width=.45\textwidth]{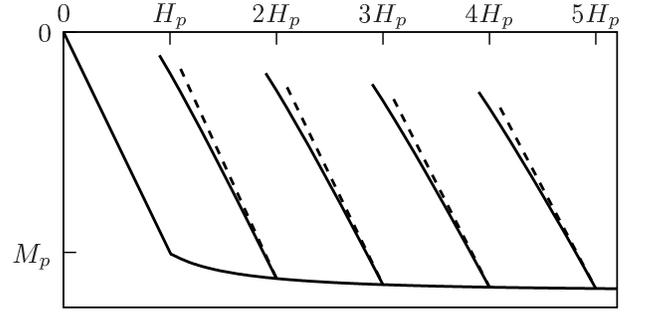}
\caption{Numerical solution for the magnetization of the descending branch for
narrow gaps $s \ll d$. The two conditions of vanishing net current and
conserved trapped flux is solved for $H^{\star} = n \hp$ (with integer $2\leq
n \leq 5$). Parameters are $w/d = d/s = 100$. The analytic result
\eqref{eq:analytic-descending-branch-2} (dashed lines) as obtained from an
expansion close to $H = H^{\star}$ gives a reasonable description of the
numerical solution over a wide field range.}
\label{fig:descending-double-strip}
\end{figure}

Although the penetration field \eqref{eq:double-penetration-field-narrow-gap} asymptotically ($s/d \to 0$) approaches the equilibrium field \eqref{eq:double-eq-field-narrow-gap}, a finite irreversibility persists and the geometric barrier rapidly reappears upon reducing the applied field. Upon decreasing
the magnetic field from a maximal value $H^{\star}$, the vortex dome expands
while keeping the trapped flux constant,
\begin{align}\label{eq:double-flux-through-dome}
   \phi_{d}(H) &\equiv \int\limits_{\bone}^{\btwo} dx\, 
                     H \sqrt{\frac{(x^{2}-\bone^{2})(\btwo^{2}-x^{2})}
                          {x^{2}(W^{2}-x^{2})}}
                = \phi_{d}^{\star},
\end{align}
where $\phi_{d}^{\star} = \phi_{d}(H^{\star})$. This constraint for the
decreasing field replaces the constraint $B_{g}=\hc$ for the increasing field.
Excluding a narrow field range $H^{\star} \simeq \hp$, the dome extends over a
large fraction of the sample and the constraint of conserved trapped flux can
be simplified under the assumptions $\bone \ll \btwo$ and $W-\btwo \ll W$ to
read
\begin{align}
   H W \Big\{1 - \frac{\nu}{4} 
        \Big[\log\Big(\frac{16}{\nu}\Big) + 1\Big] \Big\}
        &= \phi_{d}^{\star},
\end{align}
where we used $\nu(H) = 1-[\btwo(H)/W]^{2} \ll 1$ as before. Similar to the
single strip calculations [see \eq \eqref{eq:slope}], the slope of the
magnetic response $M(H) = - H \nu(H)W^{2}/4$ is given by
\begin{align}\label{eq:slope-2}
   \frac{dM}{dH} &= -\frac{W^{2}}{4} \frac{4 - \nu}{\log(16/\nu)},
\end{align}
which is numerically close to that of the Meissner phase ($-W^{2}/4$).
At the onset of the descending branch, i.e.,  $H^{\star}-H \ll
H^{\star}$, we find an analytic expression for the magnetization of the form
\begin{align}\label{eq:analytic-descending-branch-2}
   M(H) = M(H^{\star}) - \frac{H-H^{\star}}{4} W^{2} 
   \frac{4 - \nu^{\star}}{\log(16/\nu^{\star})},
\end{align}
where $\nu^{\star} = \nu(H^{\star})$ is obtained from \eq
\eqref{eq:mu-second-order-f-t-l-0}. The above expression and the result of an
exact numerical calculation of the magnetization are shown in \fig
\ref{fig:descending-double-strip}.
\subsubsection{Magnetization of the vortex dome}
\label{sec:sec:sec:magnetic-medium-3}
In the limit $s \ll d$, the diamagnetic response of the double strip is flat
and small by the factor $\sim \sqrt{d/W}$ as compared to the single strip at
$H_p$; hence, we should verify that the magnetic response of the vortex state in the flux-filled region does not substantially alter the above results.  Following again the analysis
discussed earlier in \sect \ref{sec:sec:sec:magnetic-medium-1}, the
corrections to the magnetic response
\eqref{eq:appr-mag-f-t-l} are bounded from above by the
function
\begin{align}\label{eq:magn-correction-estimated-3}
   \delta M < \frac{H}{4\pi} \frac{H_{c1}}{H_{c1}+H} 2W d
\end{align}
leading to relative corrections
\begin{align}\label{eq:relative-corrections-double-3}
   \frac{\delta M}{M} &< \frac{H}{\hc} \frac{H_{c1}}{H_{c1}+H} \frac{4}{\pi}
\end{align}
that are small as long as $H \ll \hc$. Without surface barrier, $\hc =
H_{c1}$, the corrections become of order unity only when $H \sim H_{c1}$. On
the other hand, for a large surface barrier $\hc \gg H_{c1}$, the corrections
remain small when $H \sim \hc$. We conclude, that the contribution of the
equilibrium magnetization of the Shubnikov state to the overall magnetization
of the double strip geometry (with $s \ll d \ll w$) is small and can, in most
cases, be neglected in the entire field range $H < H_{c1}$.

\section{Several strips}\label{sec:several-strips}
So far, we have given a detailed description of the single and double strip
geometries. A discussion of three coplanar strips will reveal additional
features as compared to the previous systems, and allows for a qualitative
understanding of the response of a system of a \emph{finite} number $n \geq 3$
of coplanar strips in a parallel arrangement.

The general holomorphic field for $n$ ($n \geq 1$) parallel strips arranged
symmetrically around the origin $\xi=0$ assumes the form
\begin{align}\label{eq:ant-n-strip-function}
   \mathcal{B}(\xi) &=
    H \sqrt{\prod\limits_{i}\frac{\xi^{2}-b_{i}^{2}(H)}{\xi^{2}-e_{i}^{2}}},
\end{align}
where $\pm e_{i}$ denote the strip edges and the parameters $b_{i}(H)$ define
the boundaries of the vortex states. For an even number $n = 2m$ of strips, the
$k$th strip ($0 < k \leq m$) as counted along the positive $x$-axis ranges from
$e_{2k-1}$ to $e_{2k}$ and vortices fill the region $b_{2k-1}$ to $b_{2k}$.
Every strip has a symmetric counterpart on the negative $x$-axis. For an odd
number $n=2m+1$ of strips, the above remains unchanged except for an
additional innermost strip ranging from $-e_{0}$ to $e_{0}$ with a dome
between $-\bz$ and $\bz$. The product in \eq \eqref{eq:ant-n-strip-function}
runs from 1 to $n$ (from 0 to $n-1$) for the even (odd) numbered
configurations. The expressions \eqref{eq:ant-single-strip-function} and
\eqref{eq:ant-double-strip-function} are special cases for the single and
double strip geometries. The magnetization of the $n$-strip system as
obtained from \eq \eqref{eq:magnetization-through-B} reads
\begin{align}
      \label{eq:n-strip-magnetization}
       M(H) &= -\frac{H}{4} \sum\limits_{i}(e_{i}^{2}-b_{i}^{2}).
\end{align}
In this section, we consider strips of equal width $2w$ and separated by a gap
$2s$. We also limit the analysis to the thin strip case, i.e., the thickness
$d$ of the strips is the smallest of all geometric lengths.

\subsection{Three strips}\label{sec:sec:three-strips}
The holomorphic field for three parallel strips reads
\begin{align}\label{eq:3-strip-function}
   \mathcal{B}(\xi) &= H \sqrt{
                           \frac{[\xi^{2}-\bz^{2}(H)]
                                 [\xi^{2}-\bone^{2}(H)]
                                 [\xi^{2}-\btwo^{2}(H)]}
                                {(\xi^{2}-e_{0}^{2})
                                 (\xi^{2}-e_{1}^{2})
                                 (\xi^{2}-e_{2}^{2})}},
\end{align}
with $\ez = w$, $\eo = w + 2s$, and $\et = 3w + 2s$.

\subsubsection{Meissner state}\label{sec:sec:sec:3-strip-Meissner}
For small fields, where the entire system is in the Meissner state, i.e., no
vortices have penetrated in either of the strips, the holomorphic field
reduces to
\begin{align}\label{eq:field-function-three-strips}
   \mathcal{B}(\xi) &= H\sqrt{\frac{\xi^{2} (\xi^{2}-\a^{2})^{2}}
   {(\xi^{2}-\ez^{2})(\xi^{2}-\eo^{2})(\xi^{2}-\et^{2})}},
\end{align}
where the remaining parameter $\a$ ($=\bone = \btwo$, note that $\bz = 0$) is
determined from requiring a vanishing total current in the outer strip pair,
\begin{align}\label{eq:neutral-current-3-strips}
   \int\limits_{\eo}^{\et} &\frac{dx\, x\, b^{2}}
   {\sqrt{(x^{2}-\ez^{2})(x^{2}-\eo^{2})(\et^{2}-x^{2})}} \\
   &\quad\qquad\qquad\qquad= \int\limits_{\eo}^{\et} 
   \frac{dx\, x^{3}}{\sqrt{(x^{2}-\ez^{2})(x^{2}-\eo^{2})(\et^{2}-x^{2})}}.
   \nonumber
\end{align}
With the substitution $x^{2} \to \eo^{2} + (\et^{2}-\eo^{2}) t^{2}$ the
solution can formally be expressed through
\begin{align}\label{eq:a-for-3-strips}
   \a^{2} &= \ez^{2} + (\eo^{2} - \ez^{2})\frac{\E(\kappa)}{\K(\kappa)},
\end{align}
where the elliptic integrals, defined in \eqs \eqref{eq:K} and \eqref{eq:E},
are evaluated at the imaginary argument $\kappa = \sqrt{(\et^{2}-\eo^{2})
/(\ez^{2}-\eo^{2})}$, $\kappa^{2}<0$. In two asymptotic regimes the above result
simplifies to
\begin{align}\label{eq:a-for-3-strips-limits}
   \a^{2} &\simeq \left\{\begin{aligned}
        &4(w+s)^{2} &\mathrm{for\ } s \gg w,\\
        &w^{2} \Big[1 + \frac{16}{\log{(32w/s)}}\Big] &\mathrm{for\ } s \ll w.\\
                    \end{aligned}\right.
\end{align}
The first limit ($s \gg w$) describes three almost isolated strips, while in
the latter case of nearby strips with $s \ll w$ a logarithmic dependence of
$\a$ on $s$ shows up, analogous to the expression \eqref{eq:a-over-W-ratio}
for two strips. Focusing on the regime of nearby strips $s \ll w$, we find
that the flux
\begin{align}\label{eq:flux-3-strips}
   \phi_{g} &=\int\limits_{e_{0}}^{e_{1}}\! dx\,B_{z}(x) 
            \approx H 6w \frac{\pi \sqrt{2}}{3} \frac{1}{\log{(32w/s)}}
\end{align}
passing through each of the two gaps carries a substantial fraction of the
flux $\phi_{\mathrm{b}} = H 6w$ that is blocked by the strips. The field
enhancement $\sim H\sqrt{w^{2}/sd}$ at the strip edges $\ez$ and $\eo$ [see
\eq \eqref{eq:field-function-three-strips}] is found to be parametrically
$\sim\sqrt{w/s}$ larger than at the outermost edge $\et$. A more detailed
calculation reveals, that the field strength is largest near $\ez$, followed
by a slightly lower field near $\eo$,
\begin{align}
   \frac{B_{z}(\eo-d/2)}{B_{z}(\ez+d/2)}
   = 1 - \frac{21 w^{2} - 5 \a^{2}}{\a^{2}-w^{2}}\frac{s}{4w}.
\end{align}
We conclude that the critical field $\hc$ [as discussed in
\eq~\eqref{eq:condition-penetration-field}] is first reached at the edges $\pm
\ez$ (strip index $k=0$) where the field enhancement is most pronounced. Thus,
the geometrical barrier is first suppressed in the central strip and vortices
start to penetrate the innermost strip beyond
\begin{align}\label{eq:penetration-field-3-strips}
   \hp^{k=0} &\approx \hc \sqrt{\frac{s d}{8 w^{2}}} \log(32w/s).
\end{align}
This critical field is parametrically similar to the field of first
penetration of the double strip geometry, see \eq
\eqref{eq:double-penetration-field}.

\subsubsection{Penetrated state(s)}\label{sec:sec:sec:3-strip-Penetrated}
In general, for a multiple strip geometry, the strips are not equivalent and
the penetration of vortices starts at a different field value for each strip.
The penetration sequence may depend on the geometrical setup as well as on
the boundary condition at $y \to \pm \infty$ (shunted vs. unshunted ends). In
particular, we shall compare our results to the findings by Mawatari \emph{et
al.}, who considered a system of three \emph{shunted} strips in
Ref.~\sbonlinecite{Mawatari_03}.  As the external field $H$ increases beyond
$\hp^{k=0}$, vortices populate the innermost strip ($\bz \neq 0$), while the
two other strips remain free of flux ($\bone = \btwo = \a$). The field
distribution then is given by
\begin{align}\label{eq:field-function-three-strips-penetrated}
   \mathcal{B}(\xi) &= H\sqrt{\frac{(\xi^{2}-\bz^{2}) (\xi^{2}-\a^{2})^{2}}
   {(\xi^{2}-\ez^{2})(\xi^{2}-\eo^{2})(\xi^{2}-\et^{2})}},
\end{align}
where the two parameters $\bz$ and $\a$ (now both depending on $H$) are fixed
by the constraints of critical field strength $\hc$ near the edge $\ez$ and
vanishing net current
\begin{align}\label{eq:vanishing-current-three-strips}
   \int\limits_{\eo}^{\et} dx\,I(x) = 0
\end{align}
in the outer strips. The outer strip pair will first be penetrated by
vortices only at a higher field $\hp^{k=1}$, where a critical field strength
$\hc$ is reached at the edge $\eo$. At this particular field, the requirement
that the field strength is critical at both edges $\ez$ and $\eo$ while the
outer dome has not yet developed ($\bone = \btwo = \a$), gives a relation
between $\bz$ and $\a$ of the form
\begin{align}\label{eq:condition-for-hp2}
   \bz^{2} &= w^{2}\Big[1 - \frac{8(\a^{2}-w^{2})}
   {16w^{2} + 5(\a^{2}-w^{2})}\Big].
\end{align}
Inserting this relation $\bz(\a)$ into the constraint
\eqref{eq:vanishing-current-three-strips} of vanishing net current in the
outer strip fixes the last degree of freedom $\a$ and permits to express the
second penetration field through
\begin{align}\label{eq:second-penetration-field}
   \hp^{k=1} &= \hc \sqrt{\frac{32 w^{4}sd}
   {(w^{2}-\bz(\a)^{2})(\a^{2}-w^{2})^{2}}}.
\end{align}
\begin{figure}[t]
\includegraphics[width=.482\textwidth]{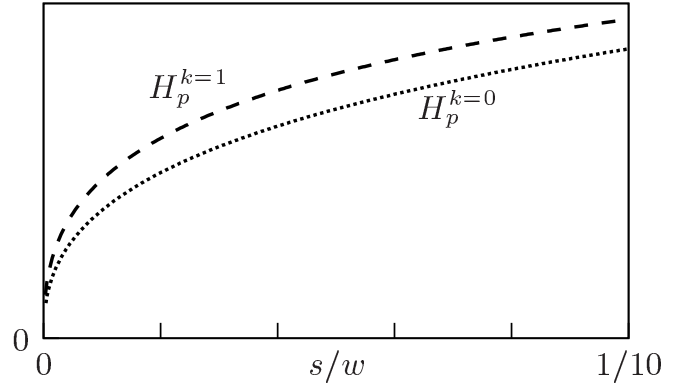}
\caption{Penetration fields $\hp^{k=0}$ (dotted) and $\hp^{k=1}$ (dashed) as a
function of $s/w$ for a system of three coplanar superconducting strips.  The
vertical scale is fixed by the specific choice of the ratio $d/w$.  The
position of the parameters $\bz$ and $\a=\bone=\btwo$ corresponding to these
two penetration fields is shown in figure \ref{fig:dome-evolution-at-hp2}.}
\label{fig:hp2}
\end{figure}%
Solving \eqs \eqref{eq:vanishing-current-three-strips} and
\eqref{eq:condition-for-hp2} numerically, we show the results for the two
penetration fields $\hp^{k=0}$ and $\hp^{k=1}$ in \fig \ref{fig:hp2}.
Using the same numerical solution, we visualize in \fig
\ref{fig:dome-evolution-at-hp2} the dome boundaries $\bz$ and $\a$ within the
strips at the first (second) penetration field $\hp^{k=0}$ ($\hp^{k=1}$) for
different values of the gap width $2s$. We observe that $\a(H)$ changes only little
between $\hp^{k=0}$ and $\hp^{k=1}$. This finding allows us to give an
estimate for $\hp^{k=1}$; indeed, inserting $\a(\hp^{k=1}) \approx
\a(\hp^{k=0})$ in \eq \eqref{eq:second-penetration-field} where
$\a(\hp^{k=0})$ is taken from \eq \eqref{eq:a-for-3-strips-limits}, we find
\begin{align}
   \frac{\hp^{k=1}}{\hp^{k=0}} &
   \approx \sqrt{\frac{w^{2}}{w^{2}-\bz(\a)^{2}}} 
   \approx \sqrt{\frac{5+\log(32w/s)}{8}}.
\end{align}
\begin{figure}[t]
\includegraphics[width=.45\textwidth]{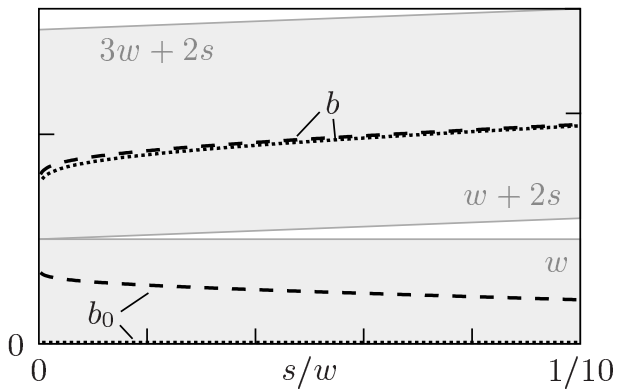}
\caption{Position of the vortex dome boundaries $\bz$ and $\a = \bone = \btwo$
of the three-strip system at the penetration fields $\hp^{k=0}$ (dotted lines)
and $\hp^{k=1}$ (dashed lines) (see also \fig \ref{fig:hp2}) as a function of
$s/w$. The strip areas along the positive $x$-axis are indicated in gray. For
$H \leq \hp^{k=0}$, the system is described by one non-vanishing parameter
$\a$ ($\bz = 0$); its dependence on $s$ is shown as a dotted line. Increasing
$H$ beyond $\hp^{k=0}$, a vortex dome forms in the innermost strip ($0< \bz <
w$) reaching a finite width at the second penetration field $\hp^{k=1}$
(dashed line). For that field, the position $\a$ has shifted towards the
center of the outer strip $2w+2s$ (dashed line). Beyond $\hp^{k=1}$, a pair of
domes forms in the two outer strips ($\bone \neq \btwo$, not shown here).}
\label{fig:dome-evolution-at-hp2}
\end{figure}%
Beyond $\hp^{k=1}$, all three strips are penetrated, and the dome widths are
determined by the restriction of no net current in the outer strips and the
two critical field conditions at the edges $\ez$ and $\eo$. Note, that the
central strip is penetrated from both edges while the strips of the pair $k=1$
are penetrated from the inner edges $\pm\eo$ only.

The order in which the strips are populated with vortices depends on the
specification of the problem. Indeed, if the same geometrical configuration
was \emph{shunted} at both ends ($y \to \pm \infty$), as considered in
Ref.~\sbonlinecite{Mawatari_03}, the outer strip pair would be populated by
vortices from $\pm e_2$, while the innermost strip remains free of
flux until much higher fields.

\subsection{Many strips - General picture}\label{sec:sec:many-strips}
For any finite number of \emph{unshunted} strips, the field enhancement in the Meissner state
is strongest at the innermost edges, i.e., the inner edges at $\pm\eo$ for an
even number of strips and the two edges at $\pm\ez$ of the central strip for
an odd number of strips (the only case where a strip is penetrated from both
sides). Subsequently, the strips are always penetrated asymmetrically from the
inner edges.  Specifically, when the applied field is raised beyond the first
penetration field, vortices start to populate the innermost strip(s) through
the respective edges, while all other strips are still free of flux. Under
further increase of $H$, the critical field strength $\hc$ is successively
reached at the inner strip edges $e_{2k-1}$ and vortices penetrate the strip
pair $k$, when $H > \hp^{k}$, with $\hp^{k} > \hp^{k-1}$, with $k$ starting
from 1 (2) in the case of an odd (even) number of strips. In the limit of a
large number of strips, the field strengths in the different gaps are almost
the same, such that vortex penetration starts within a narrow field range in
all the strips.

\section{Summary and Conclusions}\label{sec:conclusions}
In summary, we have investigated the magnetic response of (two) aligned
superconducting strips subject to a perpendicular magnetic field $H$.  The
penetration of vortices in these systems is dominated by a macroscopic energy
barrier, the so-called geometrical barrier. We have found that a narrow slit
between the strips completely suppresses this geometrical barrier as
manifested in an early field penetration and the collapse of the hysteretic
magnetization loop.  We have compared the results for a pair of rectangular
(platelet shaped) strips to those of various other shapes and geometries. Of
particular interest is the comparison with a single elliptic strip, i.e., the
generic shape defining demagnetization effects, and a single platelet strip,
the simplest system exhibiting a geometric barrier. In the elliptic case,
vortices penetrate the sample above a field $\hp = \hc d/2w$ ($H_{c1} \leq \hc
< H_{c}$), distribute uniformly inside the sample, and produce a reversible
response (in the absence of a surface barrier). The penetration of vortices in
a platelet sample is impeded by a geometrical energy barrier (and potentially
an additional surface barrier) at the sample edges, $\hp= \hc\sqrt{d/w}$. Once
this barrier is overcome, vortices occupy a finite region inside the sample
(vortex dome), while the rest carries the diamagnetic shielding currents. Upon
decreasing the applied field, the strip shows a irreversible response, where
the penetrated flux is trapped inside the sample until the vortex dome expands
to the sample edges.  These qualitative features remain valid for an array of
rectangular strips.

The attention of the present work has mainly focused on the double strip. In
the regime of a small gap parameter $s \ll w$, where the strip system is
equivalent to a single strip of width $2W = 4w + 2s$ cut in half by a narrow
gap, the geometrical barrier is overcome at much lower applied fields $H$. When
$d$ is the smallest geometric length, the situation still resembles the
one of the single strip; modifications concern the field of first penetration,
$\hp = \hc \sqrt{d/W} \boldsymbol{[}\sqrt{s/W} \log(4W/s)\boldsymbol{]}$, the
exclusive penetration from the inner edges, and the asymmetric shape of the
vortex domes leaning towards the gap. When the gap width $2s$ drops below the
thickness $d$, the currents rearrange strongly, piling up at the inner
surfaces and channeling a larger field through the central opening. In the
limit $s/d \to 0$, the geometrical barrier is maximally suppressed and the
penetration field $\hp = \hc d/2W$ of the double strip coincides with that of
a single ellipse with aspect ratio $d/2W$. In contrast to the previously
discussed cases where the magnetization decreases beyond penetration, here the
magnetic response levels off at the magnitude $M = \hc W d/2\pi$, a factor
$4/\pi$ above the magnetization at the penetration field.

In order to study the irreversibility due to the geometric barrier, we have
examined the descending branch in the magnetization upon reduction of the
field. Remarkably, our analytic results show, that the initial slope of the
descending branch is close to the Meissner response, with a correction factor
approaching unity when the reversing field approaches $\hp$ from above;
reversing the field at larger $H$, the Meissner slope is changed by a factor
$(4-\nu)/\log(16/\nu)$, where $\nu$ depends only on the point $(H,M)$ where
the slope is evaluated, $\nu = -4M/Hw^2$. Surprisingly, the correction factor
remains close to unity over a wide range of $\nu < 1$. The latter result is
equally valid for both the single and narrow-gap ($s \ll d$) double strip.

In addition, we have examined the influence of the vortex currents in the
Shubnikov phase; while our complex-analysis approach describes the
vortex-phase in terms of a non-magnetic medium, it intrinsically exhibits a
finite magnetic response.  We have shown that the magnetization due to the
structure of the vortex state in the dome remains small within the region $H <
\hc$ where our analysis is valid.  Finally, we have extended our analysis to
$n$ ($\geq 3$) strips and have given a qualitative discussion of the field
penetration for this more complex geometry.

The suppression of the geometrical barrier can be beneficial in many
circumstances, as the hysteretic behavior due to geometrical effects often
obscures other interesting physical phenomena.  E.g., this has been the case
in the identification of the vortex-lattice melting-transition in
platelet-shaped layered BiSCCO samples, where the irreversibility line
potentially interferes with the first-order melting line: polishing the sample
into a prism shape, the geometrical barrier could be suppressed, what allowed
to demonstrate experimentally that melting and irreversibility are
uncorrelated phenomena \cite{Majer_95}. Another example is the competition
between bulk pinning of vortices and pinning due to surface- and shape effects
as analyzed in the present work: again, the suppression of the geometrical
barrier provides access to an unambiguous study of bulk pinning phenomena. In
evaluating different means to suppress the geometrical barrier, the generation
of a simple gap or crack in the sample appears as a rather simple alternative.
Recently, the suppression of geometrical barriers in platelet BiSCCO samples
has also been observed when tilting the magnetic field \cite{Segev_11}.  This
finding has been related to the appearance of Josephson vortex stacks due to
the parallel field component weakening the superconductor and channeling the
perpendicular component of the magnetic field into the sample.  Relating our
present study to this experiment, we have modeled a stack of Josephson
vortices by a sample crack (of width $2s$) and observe a similar suppression
of the geometrical barrier.

Another topic where the suppression of geometrical barriers is advantageous is
the generation of low-density vortex states, which are difficult to realize in
bulk samples due to the rapid accumulation of vortices when increasing $H$
beyond $\hc$. In elliptic samples, low vortex densities of the order of
$\hp/\Phi_{0} \sim \hc (d/w)/\Phi_0$ could be achieved; however, it appears
difficult to fabricate samples with this shape.  In a realistic platelet-shape
sample, typical vortex densities are larger, of order $\hc\sqrt{d/w}
/\Phi_{0}$.  Introducing a narrow gap in the sample suppresses the geometrical
barrier and low vortex densities $\hc(d/w)\Phi_{0}$ can be reached.

Further possible applications of the narrow-gap double strip include the 
lensing of magnetic fields near the gap, what may be useful for focusing weak
magnetic signals.  Finally, the analysis and results discussed in this paper
may be of relevance in the design of superconducting atom chips for the
manipulation of ultra-cold atoms \cite{Bernon_13}.

\begin{acknowledgments}
We acknowledge illuminating discussions with Alexei Koshelev and Eli Zeldov
and the financial support of the Swiss National Fonds under the program NCCR
MaNEP.
\end{acknowledgments}


\begin{thebibliography}{34}
\bibitem[{foo({\natexlab{a}})}]{footnote:enhancement-ellipse}
\bibinfo{note}{Note that a distance $d$ away from the edges of a flat elliptic
  cylinder, the field is enhanced by the same factor $\sim\sqrt{w/d}$, while,
  upon further approach, the field increases by another factor $\sim\sqrt{w/d}$
  to a total enhancement of $\sim w/d$.}
\bibitem[{foo({\natexlab{b}})}]{footnote:finite-lambda}
\bibinfo{note}{For a finite length $\lambda \ll d$, the geometric width $2s$ is
  substituted by the real gap width $2(s + \lambda)$.}
\end{thebibliography}

\end{document}